\pdfoutput=1

\documentclass[fleqn,usenatbib]{mnras}

\usepackage{newtxtext,newtxmath}
\usepackage[T1]{fontenc}
\usepackage{ae,aecompl}
\usepackage{graphicx}
\usepackage{amsmath}
\usepackage{amssymb}
\usepackage{times}
\usepackage[dvipsnames]{xcolor}

\hypersetup{colorlinks=true,linkcolor=blue,citecolor=blue,filecolor=blue,urlcolor=blue,}

\title[The Three Causes of Low-Mass Assembly Bias]{The Three Causes of Low-Mass Assembly Bias}

\author[Mansfield \& Kravtsov]{Philip Mansfield$^{1,2}$\thanks{mansfield@uchicago.edu}, Andrey V. Kravtsov$^{1,2,3}$
\\
$^{1}${Department of Astronomy \& Astrophysics, The University of Chicago, Chicago, IL 60637 USA}\\
$^{2}${Kavli Institute for Cosmological Physics, The University of Chicago, Chicago, IL 60637 USA}\\
$^{3}${Enrico Fermi Institute, The University of Chicago, Chicago, IL 60637 USA}\\
}

\begin{document}
\label{firstpage}
\pagerange{\pageref{firstpage}--\pageref{lastpage}}
\pubyear{2018}
\maketitle

\begin{abstract}
 We present a detailed analysis of the physical processes that cause halo assembly bias -- the dependence of halo clustering on proxies of halo formation time. We focus on the origin of assembly bias in the mass range corresponding to the hosts of typical galaxies and use halo concentration as our chief proxy of halo formation time. We also repeat our key analyses across a broad range of halo masses and for alternative formation time definitions. We show that splashback subhaloes are responsible for two thirds of the assembly bias signal, but do not account for the entire effect. After splashback subhaloes have been removed, we find that the remaining assembly bias signal is due to a relatively small fraction ($\lesssim 10\%$) of haloes in dense regions. We test a number of additional physical processes thought to contribute to assembly bias and demonstrate that the two key processes are the slowing of mass growth by large-scale tidal fields and by the high velocities of ambient matter in sheets and filaments. We also rule out several other proposed physical causes of halo assembly bias. Based on our results, we argue that there are three processes that contribute to assembly bias of low-mass halos: large-scale tidal fields, gravitational heating due to the collapse of large-scale structures, and splashback subhaloes located outside the virial radius.
\end{abstract} 

\begin{keywords}
dark matter -- theory
\end{keywords}

\section{Introduction}
\label{sec:introduction}

The most visually striking feature of the large-scale structure of the universe is the clustered, web-like distribution of galaxies, with vast voids separated by walls and filaments \citep[e.g.,][]{Bond_et_al_1996}.
Understanding the clustering of galaxies within the context of the $\Lambda+$Cold Dark Matter ($\Lambda$CDM) model relies on the generic model in which galaxies are formed by the dissipation of diffuse baryon plasma within growing dark matter haloes
\citep[e.g.,][]{White_Rees_1978}. Galaxy clustering is then interpreted in terms of the clustering of dark matter haloes \citep[e.g., see][for recent reviews]{Desjacques_et_al_2018,Wechsler_Tinker_2018}, which is generally different from that of matter, i.e. the distribution of haloes is ``biased,'' relative to the mass distribution \citep{Kaiser_1984}. 

Halo bias depends primarily on halo mass
\citep[e.g.,][]{Mo_White_1996,Sheth_Tormen_1999} and this dependence is now both well-understood theoretically and well-calibrated numerically \citep{Desjacques_et_al_2018}. It is also now known that halo bias has secondary dependences on other halo properties, such as formation time, concentration, spin, and ellipticity \citep{Gao_et_al_2005,Wechsler_et_al_2006,Harker_et_al_2006,Gao_White_2007,Jing_et_al_2007,Li_et_al_2008,Faltenbacher_White_2010,Villarreal_et_al_2017,Sato-Polito_et_al_2018,Han_et_al_2018}. The first such secondary dependence was found for halo formation time and its closely related proxy -- halo concentration \citep{Gao_et_al_2005,Wechsler_et_al_2006,Harker_et_al_2006,Jing_et_al_2007} and has become known as ``assembly bias.'' Specifically, the bias of ``old'' haloes (early formation time) is generally different than that of ``young'' (late formation time) haloes, with the difference depending on halo mass and the definition of formation time\citep[e.g.,][]{Li_et_al_2008}.

Assembly bias is important for the theoretical interpretation of galaxy clustering and its potential to provide useful cosmological constraints \citep[e.g.,][]{Abazajian_et_al_2005}. There have been significant observational efforts to detect the related but distinct phenomenon of \emph{galaxy} assembly bias, the dependence of galaxy clustering on secondary halo properties, which themselves experience \emph{halo} assembly bias  \citep{Zentner_et_al_2014,Wechsler_Tinker_2018}. This is a difficult task, because the effect needs to be unambiguously disentangled from the dependence of galaxy properties on halo mass and satellite classification scheme, meaning that observational signatures of galaxy assembly bias have as yet proved to be elusive \citep{Campbell_et_al_2015,Lin_et_al_2016}. If halo assembly bias does have a signature in galaxy clustering, it would be important to understand its physical origin in order to construct robust and accurate models. Conversely, if halo assembly bias does not have observational signatures, it would be important to understand why tracers of halo age and tracers of galaxy age behave differently. We note, however, that this paper concerns itself exclusively with halo assembly bias and not with galaxy assembly bias.

The focus of this paper is to understand the physical origin of halo assembly bias, particularly in the regime of galaxy-scale halo masses. This is distinct from the origin of assembly bias at large masses, which is related to the properties of the peaks of the initial Gaussian density perturbations from which these massive haloes collapse \citep{Dalal_et_al_2008}. Peaks with the same mass but different curvature will cluster differently because peaks with larger curvatures are located in lower-density environments, while peaks with smaller curvatures are in higher-density regions. This gives rise to assembly bias because peak curvature is directly related to a halo's mass accretion history, which is also affected by tidal torques from the surrounding anisotropic mass distribution \citep{Desjacques_2008}. Although this curvature-related bias can be
reduced by compensating effects found in some proxies of halo age \cite[][]{Zentner_2007,Sandvik_et_al_2007,Mao_et_al_2018}, it is present for other age definitions and when more physical definitions of halo boundaries and masses are used \citep{Chue_et_al_2018}. 

At smaller halo masses, however, the physics of assembly bias is more complex because the mass evolution of haloes is determined by a combination of the properties of their initial density peaks, and also by non-linear processes \citep[e.g.,][]{Wang_et_al_2007,Hahn_et_al_2009}. The simple and striking manifestation of this is that the sign of assembly bias switches for small-mass haloes when $c_{\rm vir}$ is used as a measure of halo age \citep{Wechsler_et_al_2006,Dalal_et_al_2008}.

A number of studies have explored the physical processes that can give rise to halo assembly bias in the small-mass regime. One readily apparent process is the non-linear effects that a massive host halo can exert on its smaller-mass neighbours.
In particular, ``splashback'' (often also called ``backsplash'') subhaloes pass within the inner regions of a larger halo but are located outside its virial radius at the epoch of analysis. Such haloes appear isolated, but will have had their mass accretion histories truncated due to their previous close encounters with their hosts and have thus been studied as a potential source of low-mass assembly bias \citep{Wang_et_al_2009,Li_et_al_2013,Wetzel_et_al_2014,Sunayama_et_al_2016}. 

\begin{figure*}
   \centering
   \includegraphics[width=2\columnwidth]{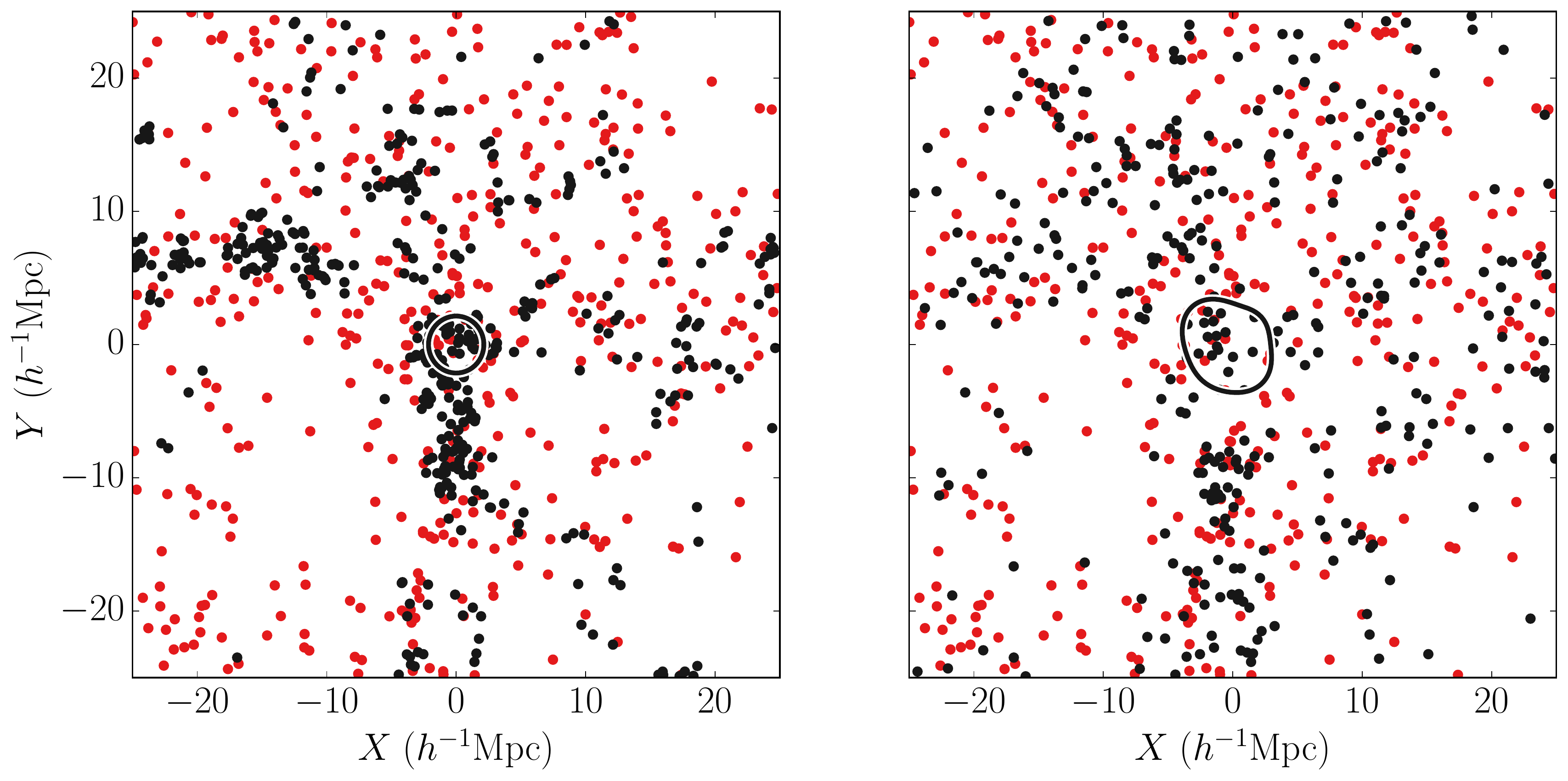}
   \caption{The effect of subhalo classification on the apparent distribution of ``old'' and ``young'' haloes and their relative clustering. Both panels show the distribution of haloes in a 25 $h^{-1}$Mpc cube around the largest cluster in the Bolshoi simulation. In the left panel, haloes within the virial radii of larger hosts have been classified as subhaloes and removed. The 15\% of haloes with the smallest $c_{\rm vir}$ (``young'' haloes)  are plotted in red, while the 15\% of haloes with the largest $c_{\rm vir}$ (``old'' haloes) are plotted in black. For scale, the virial radius of the central cluster is shown as a black circle. While both young and old halo samples are distributed non-uniformly, old haloes cluster more strongly and form prominent structures on scales exceeding $\approx 20\ h^{-1}$Mpc.
            Right panel: the same volume but all halos within the splashback shell of a larger host have been classified as subhaloes and removed. For scale, the splashback shell of the central cluster is plotted in black.
            The age-dependent clustering of haloes in the right panel, while still visually apparent, is significantly weaker. This is because splashback subhaloes are preferentially old and trace the more clustered distribution of their massive hosts.}
   \label{fig:visual_clustering}
\end{figure*}

Although splashback subhaloes are mostly found within three virial radii of their host halo,
they can give rise to an assembly bias signal at much larger distances. This is because at large scales the spatial distribution of splashback subhaloes will track the distribution of their massive hosts and will therefore be more strongly clustered than that of distinct haloes. A similar effect would occur if subhaloes located within the virial radius of their host were included in the sample used to measure halo clustering and assembly bias. This is illustrated in Fig.~\ref{fig:visual_clustering}, which compares the clustering of early- and late-forming haloes with splashback subhaloes included and removed, respectively. Removing splashback subhaloes significantly reduces the difference in clustering between the two halo samples, even on scales much larger than the virial radius of the most massive haloes within the volume.
Nevertheless,  multiple studies have demonstrated that splashback subhaloes alone cannot be responsible for the entire assembly bias signal \citep{Wang_et_al_2009,Sunayama_et_al_2016}, a fact that can be seen visually in Fig.~\ref{fig:visual_clustering}. A similar conclusion was reached by \citet{Hearin_et_al_2015}, albeit in the related but distinct context of galactic conformity.

Another process that could contribute to assembly bias is the truncation of a halo's mass growth by the tidal force generated its most gravitationally-dominant neighbor \citep{Hahn_et_al_2009,Behroozi_et_al_2014,Hearin_et_al_2016_2,Salcedo_et_al_2018}. Even though single-halo tidal forces become small beyond $\approx 3\--5 \times R_{\rm vir}$ of the host, the haloes truncated by these forces can give rise to large-scale assembly bias in a way similar to splashback subhaloes. A similar truncation of halo mass growth can be caused by the overall tidal force from all of the surrounding haloes and structures in the matter distribution \citep{Hahn_et_al_2009,Wang_et_al_2011,Paranjape_et_al_2018,Musso_et_al_2018}, as the largest filaments and sheets generate strong tidal fields throughout their volumes. Since these structures can be several tens of Mpc in size, they can comfortably give rise to assembly bias on large scales. This effect has been characterized in terms of both the tidal force and the anisotropy of the tidal field, although, in practice, a high degree of anisotropy tends to correlate with the magnitude of the tidal force, so it is not clear that the two effects can be separated cleanly.

Finally, the gravitational heating of matter  within large-scale structure structure has been proposed as a process that can contribute to assembly bias \citep{Wang_et_al_2007,Dalal_et_al_2008}. For example, matter within the deep potentials of filaments can acquire large velocities during accretion, and thus cannot be accreted by small-mass haloes located within the filament. The smaller accretion rates of such haloes would thus give rise to assembly bias. Note that although gravitational heating and strong tidal forces generally happen in similar regions, they are physically distinct phenomena: tidal forces arise via large second derivatives in the gravitational potential, while gravitational heating is caused by the potential depth.

Although significant effort has been devoted to studying these effects
\citep{Wang_et_al_2007,Dalal_et_al_2008,Wang_et_al_2009,Hahn_et_al_2009,Wang_et_al_2011,Li_et_al_2013,Wetzel_et_al_2014,Sunayama_et_al_2016,Hearin_et_al_2016_2,Paranjape_et_al_2018,Salcedo_et_al_2018,Musso_et_al_2018}, their relative importance and  a coherent physical picture for the origin of low-mass halo assembly bias has not yet been established. The primary goal of this paper is to rectify this. To this end, we define a set of quantitative proxies for each of the different processes outlined above and use them to investigate the relative contribution of these processes to the low-mass assembly bias signal. Specifically, we examine how efficiently sample cuts defined by each proxy can remove the signal.

The paper is organized as follows. In section \ref{sec:methods} we describe basic definitions and measurements and describe our cosmological simulations and halo sample, with sections \ref{ssec:measure_tidal}-\ref{ssec:grav_heating} focusing on the proxies of the processes described above, and section \ref{ssec:connection} describing the core methodology of this paper. In section \ref{sec:analysis} we present measurements and estimates of the relative contribution of different processes to low-mass assembly bias. We discuss topics related to the interpretation of this work in section \ref{sec:discussion} and summarize our results in \ref{sec:summary}. {\it The key results of this study are presented in Fig.~\ref{fig:percent_comp}.}

\section{Methods}
\label{sec:methods}

\subsection{Simulations and codes}
\label{ssec:simulations_and_codes}

In this paper we use halo catalogues and particle snapshots from the Bolshoi and BolshoiP cosmological dark matter-only $N$-body simulations, made available through the CosmoSim\footnote{\href{https://www.cosmosim.org}{https://www.cosmosim.org}} and Skies and Universes\footnote{\href{http://skiesanduniverses.iaa.es}{http://skiesanduniverses.iaa.es}} databases \citep{Klypin_et_al_2017}. Both simulations followed the evolution of 2048$^3$ particles in boxes of size 250 $h^{-1}$ Mpc using the ART code, with force resolutions of $\approx 1\,h^{-1}$ kpc.
The Bolshoi simulation assumed $\Lambda$CDM cosmology with parameters consistent with the WMAP 9 year constraints: $\Omega_{\rm M} = 0.270,$ $\Omega_{\rm B}=0.047,$ $\sigma_8=0.820,$ $n_s=0.95,$ and $H_0 = 70.0\ {\rm km\ s^{-1}\,Mpc^{-1}}$. The BolshoiP simulations assumed parameters consistent with the constraints from the Planck observatory: 
$\Omega_{\rm M} = 0.307,$ $\Omega_{\rm B}=0.048,$ $\sigma_8=0.823,$ $n_s=0.96,$ and $H_0 = 67.8\ {\rm km\ s^{-1}\,Mpc^{-1}}$. The corresponding particle masses are $1.3\times 10^8\ h^{-1}\,M_\odot$ and $1.5\times 10^8\ h^{-1}\,M_\odot$, respectively. Additional technical information, such as time stepping criteria and initial conditions, can be found in \citet{Klypin_et_al_2011} and \citet{Klypin_et_al_2016}. Although all plots shown in this paper use Bolshoi data, all analyses were repeated for BolshoiP with similar results. We also use the simulation suite described in \citet{Diemer_Kravtsov_2015} for some convergence and performance testing.

Haloes in the Bolshoi and BolshoiP simulations were identified using version 0.99RC2+ of the \textsc{Rockstar} halo finder \citep{Behroozi_et_al_2013_1}, and version 1.0+ of the related consistent-trees method \citep{Behroozi_et_al_2013_2} was used to construct halo merger trees. The catalogues and merger trees we use were downloaded from the CosmoSim database. We use the \textsc{Shellfish} algorithm to identify splashback shells -- the 3D surfaces formed by the outermost apocentres of accreted matter \citep{Mansfield_et_al_2017}\footnote{\href{https://github.com/phil-mansfield/shellfish}{https://github.com/phil-mansfield/shellfish}}. We use the {\tt Colossus} python package \citep{Diemer_2017_2}\footnote{\href{http://www.benediktdiemer.com/code/colossus/}{http://www.benediktdiemer.com/code/colossus/}} to calculate various relevant cosmological quantities and statistics  and the halotools package \citep{Hearin_et_al_2016}\footnote{\href{https://halotools.readthedocs.io}{https://halotools.readthedocs.io}} to calculate correlation functions efficiently .

\subsection{Basic halo properties}
\label{ssec:halo_properties}

Throughout this paper, we use halo masses, $M_\Delta,$ defined within a radius, $R_\Delta$, enclosing a specified density contrast $\Delta=\rho(< R_\Delta)/\bar{\rho}_{\rm m}$:
\begin{equation}
	\label{eq:m_delta}
   M_\Delta = \frac{4\pi}{3}\Delta\,\bar{\rho}_{\rm m}\, R_\Delta^3,
\end{equation}
where $\bar{\rho}_{\rm m}$ is the mean density of matter. 
We use the ``virial'' density contrast defined as 
\begin{align}
	\label{eq:delta_vir}
   \Delta = \frac{18\pi^2 + 82\widetilde{x} - 39\widetilde{x}^2}{\Omega_{\rm M}},
\end{align}
where $\widetilde{x} = \Omega_{\rm M}(z) - 1$. This definition formally corresponds to the virialization density of a spherical tophat perturbation collapsing in a flat $\Lambda$CDM universe at the redshift of analysis \citep{Bryan_Norman_1998}.  This choice is dictated
by the fact that \textsc{Rockstar} calculates properties of haloes using particles contained within friends-of-friends (FOF) groups identified using the linking length of $b=0.28$ (in units of mean interparticle separation) in the Bolshoi and BolshoiP halo catalogues. Such linking length ensures full percolation only out to density contrasts $\gtrsim\Delta_{\rm vir}$ \citep{Behroozi_et_al_2013_1}, which means that not all particles are included in the groups at distances corresponding to lower $\Delta$. This, in turn, biases halo properies such as concentration.   Additionally, density profiles of haloes begin to deviate from the NFW (Navarro–Frenk–White) form beyond $R_{\rm vir}$ \citep[e.g.,][]{Becker_Kravtsov_2011,Diemer_Kravtsov_2014}, which means that NFW profile fits at $R\gtrsim R_{\rm vir}$ are generally biased (see Appendix \ref{app:halo_definition_rockstar} for an extended discussion on this issue).

We note, however, that $R_{\rm vir}$ does not correspond to an actual \emph{physical} boundary of the halo (for detailed discussion see, e.g., \citealp{More_et_al_2015} as well as section \ref{ssec:subhaloes} below). As such, we would have been free to select many other equally valid choices of $\Delta$ for our baseline definition, so long as they are cross-matched with profile properties measured from a catalogue which uses $\Delta \geq \Delta_{\rm vir}$. We discuss the significance of this choice when interpreting our results in section \ref{ssec:definition_dependence} and find that it is not a significant issue.

We assume that the radial density profiles of dark matter haloes are well-approximated by an NFW profile within $R_{\rm vir}$:
\begin{equation}
	\label{eq:nfw}
   \rho(r) = \frac{\rho_s}{x(1 + x)^2},
\end{equation}
where $x \equiv r/R_{\rm s}$ \citep{Navarro_et_al_1997}. This profile is fully specified by a combination of $R_{\rm vir}$ and the ``concentration,'' $c_{\rm vir}=R_{\rm vir}/R_{\rm s}$. We use the $c_{\rm vir}$ values measured by the \textsc{Rockstar} code, which calculates these values by breaking particles within $R_{\rm vir}$ into radial bins containing at least 15 particles, calculating densities in each bin, and performing a $\chi^2$-minimization fit to Eq.~\ref{eq:nfw}

We also make use of halo circular velocity profiles, $V(r) = \sqrt{GM(<r)/r}$, and its maximum values, $V_{\rm max}$. The largest value of $V_{\rm max}$ during the evolution of a halo is denoted as $V_{\rm peak}$. $V_{\rm max}$ and $V_{\rm peak}$ are useful quantities because they allow us to define halo samples by potential depth without any dependence on our choice of halo boundary. Additionally, some halo samples will contain subhaloes, and the high ambient density around these objects can mean that $R_{\rm vir}$ is not a meaningful property. $V_{\rm peak}$ is useful in particular because it is a proxy for halo potential depth prior to mass loss or tidal stripping  \citep[e.g.,][]{Nagai_Kravtsov_2005}, physical processes which both play significant roles in our analysis. Furthermore, models of galaxy clustering indicate that ``peak'' mass definitions are better correlated with the observable properties of galaxies than values at the current epoch \citep{Reddick_et_al_2013}.

There are a number of definitions of halo age used in the literature: single-epoch accretion rates \citep[e.g.,][]{Lacey_Cole_1993,Li_et_al_2008}, current halo properties - such as concentration - related to a halo's mass accretion history \citep[e.g.,][]{Wechsler_et_al_2006,Villarreal_et_al_2017,Sato-Polito_et_al_2018}, the epoch at which a halo first achieved half of its current mass \citep[e.g.,][]{Gao_et_al_2005}, or a characteristic timescale of an analytic fit to halo mass accretion history \citep[e.g.,][]{Wechsler_et_al_2002,Zentner_2007}. In this paper, we primarily adopt $c_{\rm vir}$ as a tracer of halo age, with older haloes having larger concentrations. We briefly explore the effect of using different definitions in section \ref{ssec:definition_dependence}.

We focus on $c_{\rm vir}$ for several reasons. First, $c_{\rm vir}$ has been  demonstrated to strongly correlate with a number of explicit indicators of halo age \citep{Bullock_et_al_2001,Wechsler_et_al_2002,Zhao_et_al_2003,Lu_et_al_2006,Ludlow_et_al_2013,Ludlow_et_al_2014}. Second, the connection between accretion history and $c_{\rm vir}$ has a solid theoretical underpinning \citep{Zhao_et_al_2003,Lu_et_al_2006,Dalal_et_al_2010}, as demonstrated by the accuracy of the concentration models based on halo mass accretion history \citep[e.g.][]{Zhao_et_al_2003,Zhao_et_al_2009,Dalal_et_al_2010,Ludlow_et_al_2014,Diemer_Joyce_2018}. Third, the convergence criteria for halo density profiles  \citep[e.g.][]{Power_et_al_2003,Navarro_et_al_2004,Springel_et_al_2008}, and for concentrations \citep[see, e.g., Section 3.2 in][]{Diemer_Kravtsov_2015} are well studied and it is thus relatively straightforward to identify regimes in which numerical concentrations can be trusted.

\subsection{Definition of halo boundaries and subhaloes}
\label{ssec:subhaloes}

Throughout this paper, we define subhaloes as the haloes located within the boundary of a larger ``host halo,'' and refer to all non-subhaloes as ``distinct haloes.'' Of course, this classification depends on the definition of halo boundary and will have a clear qualitative meaning only if we use halo boundary definition that corresponds to an actual physical boundary. 

Traditionally, spheres of radius $R_{\rm vir}$ (or some other overdensity radius) are used as halo boundaries, but this choice has a number of issues \citep[see, e.g.,][]{Diemer_et_al_2013,More_et_al_2015}. The first issue is that that there is no commonly-used overdensity radius that corresponds to any physical change or feature in the radial profiles of various halo properties \citep[see, e.g., fig.~3 of][]{Diemer_et_al_2013b}. The second issue is that many studies have established that a substantial fraction of bound subhaloes and matter have first orbits whose apocentres take objects out to as far as $\approx2-3\times R_{\rm vir}$ of the host halo \citep{Gill_et_al_2005,Ludlow_et_al_2012,Mansfield_et_al_2017,Diemer_2017_1}. 

Fortunately, haloes \emph{do} have unambiguous edges manifested as sharp drops in density and caused by the pileup of particles at the apocentres of their first orbits. These edges in form 3D surfaces called ``splashback shells,'' and enclose almost all matter and subhaloes ever accreted by a halo. Haloes outside $R_{\rm vir}$ of their host, but within its splashback shell are called ``splashback subhaloes.''\footnote{The terminology used to refer to these objects is varied: different authors refer to them as ``backsplash suhaloes'' or ``splashback subhaloes,'' and often refer to them as ``haloes'' instead of ``subhaloes.'' All these terms refer to the same concept. Some authors may use the term ``flyby [sub]haloes'' interchangeably with ``splashback [sub]haloes,'' although the former term generally implies that merger tree analysis has been used.}

Splashback subhaloes can be identified and removed in one of two ways. The first is a classification based on the past halo trajectories, where merger trees are used to determine whether a halo has ever been within a larger host \citep[e.g.,][]{Ludlow_et_al_2009,Wang_et_al_2009,Diemer_2017_1,Diemer_et_al_2017}. The second is to directly identify splashback shells of haloes and flag all haloes wthin them as subhaloes. We adopt the second approach as our fiducial classification method, but employ both throughout the paper to ensure that our results are robust and do not rely on the specifics of either approach.

For lexical clarity, we refer to subhaloes identified through merger tree analysis as ``flyby subhaloes'' and subhaloes identified through the construction of splashback shells as ``splashback subhaloes.''

\subsubsection{Flyby subhaloes}
\label{ssec:dynamical_splashback}

To identify flyby subhaloes, we use the following procedure for each halo in the $z=0$ halo catalogue. First, using consistent-trees \citep{Behroozi_et_al_2013_2}, we identify the main-line branch for the halo, labelling the $z=0$ halo the ``root halo'' and all other haloes on the branch its ``progenitor haloes.'' If any haloes on the branch are within the virial radius of another halo at any redshift, the root is flagged as a flyby subhalo.

This process is complicated by the fact that during major mergers the virial radii of both merging haloes fluctuate significantly and it is common for both host haloes to be at least temporarily identified as subhaloes of one another. This can lead to the final host halo being misidentified as a flyby subhalo of an object that no longer exists once the merger is complete. To rectify this, if the search of a root halo's progenitors reveals that some progenitor, $P,$ is within $R_{\rm vir}$ of a host halo, $H,$ we only classify the root halo of $P$ as a flyby subhalo when the following three conditions are met:
\begin{enumerate}
   \item $H$ must have a root halo at $z=0$.
   \item The root halo of $H$ must not be within $R_{\rm vir}$ of the root halo of $P.$
   \item The root halo of $H$ must have a strictly larger mass than the root halo of $P.$
\end{enumerate}
Our tests indicate that just enforcing conditions 1 and 2 is sufficient to correct the overwhelming majority of false classifications. This procedure can be extended to root redshifts other than $z=0$.

Although the identification of flyby subhaloes is well-defined and only requires the use of a merger tree, it is not without drawbacks.
First, the method uses $R_{\rm vir}$, which as we discussed above does not correspond to a physical halo boundary. Second, this approach does not distinguish between ordinary subhaloes with apocentres outside $R_{\rm vir}$, and subhaloes that may have undergone dynamical three body interactions that resulted in their unbinding and ejection and are a qualitatively distinct population from splashback subhaloes. Although a substantial fraction of subhaloes may have undergone such interactions \citep{Sales_et_al_2007,Ludlow_et_al_2009}, we find that haloes which have been ejected from the splashback shell are rare and do not have an impact on our analysis (see section  \ref{ssec:consitency_of_splashback}). Third, this method does not count haloes within the splashback shell on their first infall as subhaloes, even though this population is similar to first-infall haloes within $R_{\rm vir},$ which this method does classify as subhaloes.

\subsubsection{Splashback shell subhaloes}
\label{ssec:splashback_shells}

The simplest way to estimate the size of a halo's splashback shell it to approximate it as a sphere and estimate its radius from the location of sharp steepening it causes in the halo's density and subhalo number density profiles \citep[e.g.][]{Fillmore_Goldreich_1984,Bertschinger_1985,Diemer_Kravtsov_2014,Adhikari_et_al_2014,More_et_al_2015,Diemer_2017_1,Diemer_et_al_2017}. This radius is then called the ``splashback radius,'' $R_{\rm sp}.$ However, the application of this method for individual haloes is not straightforward \citep[see][]{Mansfield_et_al_2017}. In addition, actual splashback shells are not spherical and spherical approximation may result in misclassification of a certain fraction of subhaloes. For this reason, we use the \textsc{Shellfish} algorithm \citep{Mansfield_et_al_2017} to identify fully 3D splashback shells.

The \textsc{Shellfish} algorithm identifies splashback shells by measuring sharp density drops in many 1D density profiles along tens of thousands of lines of sight around a halo and fits a flexible smooth 3D surface  to their location \citep{Mansfield_et_al_2017}.  Once \textsc{Shellfish} has identified splashback shells, we use the efficient intersection-checking method described in Appendix \ref{app:fast_halo_containment_checks} to flag all haloes within the splashback shell of any larger halo as splashback subhaloes.

There are three complications to using \textsc{Shellfish} which must be addressed before it can be used to construct subhalo catalogs: its $N_{\rm 200m}$ convergence limit, the occurrence of rare but catastrophic fitting failures, and its behavior for low-accreting hosts. We perform extensive tests on all three issues and find that once accounted for in the ways described below, they do not have a significant effect on our results.

First, \textsc{Shellfish} has a rather stringent convergence limit and requires that haloes have more than $5\times 10^4$ particles within $R_{\rm 200m}$, the overdensity radius corresponding to $\Delta=200\,\rho_{\rm m}$, to achieve $R_{\rm sp}$ measurements with accuracy better than $5\%$. This corresponds to the $M_{\rm 200m} \gtrsim 7-8\times 10^{12}\ h^{-1}M_\odot$ or $V_{\rm peak} \gtrsim 280\ {\rm km\ s^{-1}}$ in the Bolshoi and BolshoiP boxes. Below this mass, we use the fitting formula for the median $R_{\rm sp}$ provided in \citet{Mansfield_et_al_2017}, and flag haloes within spheres of radius $R_{\rm sp}$ instead. 
Tests using the higher resolution L0063\_CBol box from \citet{Diemer_Kravtsov_2014} indicate that this results in a negligible number of subhalo misclassifications compared to using real \textsc{Shellfish}-identified splashback shells because the majority of the splashback subhaloes in our mass range have hosts larger than 280 km s$^{-1}$.

The second complication is that for a small number of host haloes \citep[$\approx 1\%$,][]{Mansfield_et_al_2017}, irregularities in the local density field cause  \textsc{Shellfish}
to fail to identify the correct surface shape, adopting a barbell-shaped surface instead, which can cause subhaloes well within $R_{\rm vir}$ to be misclassified as distinct haloes. To mitigate this, we mark haloes as splashback subhaloes if they fall within either their host's splashback shell or within a sphere centred on that host of radius $R_{\rm vir}.$ We analysed the distribution of the minimum radii of \textsc{Shellfish} shells in haloes which were visually-identified to be unaffected by this surface fitting failure and found that the minimum radii are generally larger than $R_{\rm vir}$. Thus, the procedure we adopt is unlikely to result in misclassification of distinct host haloes as subhaloes.

The third complication is that the \textsc{Shellfish} algorithm underestimates the size of splashback shells for haloes that are accreting slower than the baseline pseudo-evolution accretion rate \citep{Mansfield_et_al_2017}. However, this only lowers the splashback radius by $\approx 10\%$ and few haloes massive enough to host subhaloes in our target mass range accrete this slowly, so it is not expected to be a significant issue. Empirically, we find that virtually all flyby splashback subhaloes whose hosts are in this accretion regime are also within the splashback shells of their hosts (see section \ref{ssec:consitency_of_splashback}), so we do not explicitly account for this effect.

\subsection{Halo sample}
\label{ssec:halo_sample}

Although we will examine the mass-dependence of assembly bias in section \ref{ssec:time_evolution_of_assembly_bias}, the majority of our analysis focuses specifically on low-mass haloes. Our primary concern when defining a halo sample is to prevent the inclusion of haloes whose convergence radii are large enough that they introduce numerical effects into \textsc{Rockstar}'s measurements of $c_{\rm vir}.$ As mentioned above, the numerical reliability of density profiles has been well studied, but for cosmological simulations with small softening scales the exact convergence properties are covariant with particle count, softening scale, halo mass, and time stepping scheme \citep{Power_et_al_2003,Ludlow_et_al_2018}, so determining convergence limits for an individual simulation should always be done through the comparison of carefully constructed multi-box suites.

Because there is only a single Bolshoi box, we place an upper bound on the convergence limit using the CBol simulation suite described in \citet{Diemer_Kravtsov_2015}. Of particular note is the box CBol\_L0125, which has the same particle mass to Bolshoi, but which has much larger timesteps within halo centers, implying that the convergence radius of Bolshoi should be smaller than that of CBol\_L0125.\footnote{The difference in softening scale between these boxes makes an exact comparison difficult without a detailed analysis beyond the scope of this paper. See \citet{Diemer_Joyce_2018} for some additional discussion on the subtleties of comparing Bolshoi to this simulation suite.} We find that when using the same \textsc{Rockstar} version and configuration variables as our Bolshoi catalogue, the $V_{\rm peak}-c_{\rm vir}$ relation for the CBol\_L0125 box agrees with the higher resolution CBol\_L0063 box above $V_{\rm peak} = 120$ km s$^{-1},$ corresponding to a somewhat conservative cutoff particle count of $N_{\rm peak}\approx1.3\times 10^3$. 

Our low-mass halo sample includes haloes with $120\ {\rm km\ s^{-1}} < V_{\rm peak} < 220\ {\rm km\ s^{-1}}$ (approximately $1.7\times 10^{11}\ h^{-1}\ M_\odot < M_{\rm peak} < 1.2 \times 10^{12}\ h^{-1}\ M_\odot$). Due to the slope of the halo mass function, the majority of haloes will be close to the lower mass limit, making the choice in upper mass limit less important. We chose the upper mass limit so our sample spans roughly a factor of eight in $M_{\rm peak}$ and find that our results are not particularly sensitive to this choice.

\subsection{Measuring tidal force strength}
\label{ssec:measure_tidal}

Tidal forces have been proposed as a potential cause of assembly bias \citep{Hahn_et_al_2009,Wang_et_al_2009,Hearin_et_al_2016_2,Salcedo_et_al_2018,Paranjape_et_al_2018} because they can slow down, stop, or reverse mass accretion. These fields are strongest in dense environments, such as within large-scale filaments or near the outskirts of massive haloes, allowing distant haloes in similar environments to have correlated accretion histories. Below, we describe methods for measuring the strength of both the single-halo tidal field, and the large-scale tidal field.

\subsubsection{Tidal force from a single halo}
\label{ssec:singal_halo_tidal_fields}

A typical simplifying assumption when calculating the tidal force felt by a halo is to assume that it is primarily caused by a single massive halo. If one also assumes that the point of interest is orbiting around that halo on a circular orbit, one can compute the tidal Hill radius, $R_{\rm Hill},$ corresponding to the distance to the nearest two Lagrangian points when the effective potential is approximated to second order. However, the assumptions that are made in calculating $R_{\rm Hill}$ are not correct for distinct haloes in a $\Lambda$CDM cosmology. This is because these haloes are almost never on circular orbits around each other and, as we discuss in section \ref{ssec:r_tidal}, the tidal force generally has a significant contribution from multiple haloes and from the large-scale matter distribution. 
Thus, formally, the Hill radius is not a physically meaningful quantity for distinct haloes. Nevertheless, the classical Hill radius can be used to estimate the tidal force of a halo's most gravitationally-dominant neighbour.

As a simple and definitionally robust proxy for $R_{\rm Hill}$ we use the virial radius-scaled distance, $D_{{\rm vir},i}$, for every distinct halo $i$:
\begin{eqnarray}
    D_{{\rm vir},i} &=& {\rm min}_j \left\{\frac{R_{ij}}{R_{{\rm vir},j}}\right\}\label{eq:Dvir_def}\\
    &=& 3^{1/3}\ {\rm min}_j \left\{ R_{ij}\left(\frac{M_{{\rm vir},i}}{3M_{{\rm vir},j}}\right)^{1/3}\right\} = 3^{1/3}\ \frac{R_{{\rm Hill},i}}{R_{{\rm vir},i}}\label{eq:Dvir_Hill}
\end{eqnarray}
where $j$ runs over all distinct haloes within some search radius, $R_0,$ which are more massive than the halo, and $R_{ij}$ is the distance between haloes $i$ and $j$. 
Haloes with smaller $D_{\rm vir}$ experience larger tidal forces and haloes with larger $D_{\rm vir}$ have smaller tidal forces. As Eq.~\ref{eq:Dvir_Hill} shows, $D_{\rm vir}$ is proportional to $R_{\rm Hill}$. This means that a rank-ordering by $D_{\rm vir}$ is equivalent to a rank-ordering by $R_{\rm hill}/R_{\rm vir}$, while formally $D_{\rm vir}$ is always a well-defined quantity and also allows for easy comparison with other assembly bias studies \citep[e.g.,][]{Villarreal_et_al_2017,Salcedo_et_al_2018}

Our tests indicate that $D_{\rm vir}$ is well-converged for haloes in the mass range $120\ {\rm km\ s}^{-1}<V_{\rm peak}<220\ {\rm km\ s}^{-1}$ for $R_0 \approx 100\,R_{\rm vir}.$

\subsubsection{Large-scale tidal radius and mass}
\label{ssec:r_tidal}

Although the single-source approximation is reasonably accurate for subhaloes, our tests indicate that most distinct haloes have multiple neighbors which contribute significantly to the tidal forces they feel. Moreover, we found that large-scale structures in mass distribution, such as filaments can contribute to the tidal force experienced by haloes substantially. For example, by combining the assumption of cylindrical symmetry with the radial density profiles of filaments reported in \citet{Cautun_et_al_2014}, we construct a toy model for filament potentials. Applying this model, we find that even in moderate-sized filaments with $R_{\rm filament} \gtrsim 3\,h^{-1}\,{\rm Mpc}$, the tidal force generated by the filament is comparable to or stronger than the typical tidal force generated by a halo's single most gravitationally dominant neighbor.

For this reason, we compute the tidal radius of a halo calculated from the overall matter distribution around a halo, $R_{\rm tidal}$, as a proxy for the combined tidal force from all neighbour haloes and structures. To compute $R_{\rm tidal}$, we first construct the tidal tensor, $T_{ij}$, the Hessian of the external potential:
\begin{align}
\begin{split}
    \label{eq:tidal_tensor}
    T_{ij} = \sum_k &\frac{m_k}{(x_k^2 + y_k^2 + z_k^2)^{5/2}}\times\\
    &\begin{pmatrix} 
         y_k^2 + z_k^2 - 2x_k^2 & -3x_ky_k         & -3x_kz_k \\
        -3x_ky_k          & x_k^2 + z_k^2 - 2y_k^2 & -3y_kz_k \\
        -3x_kz_k          & -3y_kz_k               & x_k^2 + y_k^2 - 2z_k^2
    \end{pmatrix}.
\end{split}
\end{align}
Here, $k$ runs over all particles between two search radii, $R_{\rm min}$ and $R_{\rm max},$ $m_k$ is the mass of particle $k,$ and $x_k,$ $y_k,$ and $z_k$ are the components of the displacement vector from the halo centre to particle $k.$ The tidal radius lies along the steepest repulsive axis of the tidal field, and since the tidal tensor, like all Hessians, equivalently describes the second derivatives at the origin of a paraboloid with eigenvectors pointing along the paraboloid's axes, the tidal field along the chief repulsive axis is given by
\begin{equation}
    \Phi_{\rm steepest}(r_1) = \frac{1}{2}\alpha_1 r_1^2,
\end{equation}
where $\alpha_1$ is the most negative eigenvalue of $T_{ij},$ and $r_1$ is the radial distance along the corresponding eigenvector. We then assume that all non-tidal pseudo-forces (most notably the centrifugal force) are small and that at large distances the halo's mass is well-approximated by $M_{\rm vir}$, making the tidal radius and the corresponding tidal mass
\begin{equation}
	\label{eq:r_tidal}
    R_{\rm tidal} = \left(\frac{GM_{\rm vir}}{\alpha_1}\right)^{1/3};\ \ \ \ M_{\rm tidal}=M(<R_{\rm tidal})
\end{equation}

To increase computational efficiency, we make two further approximations. First, we do not add the tidal contribution from any particles further than $100\,R_{\rm vir},$ and second, we subsample particles by a factor of 64 and multiply $m_k$ by 64 in Equation \ref{eq:tidal_tensor}. Our tests indicate that the combined effects of both these approximations on $R_{\rm tidal}$ are at the sub-percent level. We set $R_{\rm min} = 10\,R_{\rm vir}.$ This choice is discussed in detail in Appendix \ref{app:tidal_forces}.

Some authors have suggested that the primary feature of interest in the tidal field is its anisotropy, which can be defined in a number of ways \citep{Wang_et_al_2011,Paranjape_et_al_2018}. We chose to use $R_{\rm tidal}$ as a proxy for the total tidal force for two reasons. First, there are a number of different proxies for anisotropy and it is not clear a priori  which definition is optimal. Second, we carried out analysis of assembly bias described in sections \ref{ssec:assembly_bias_metrics} and \ref{ssec:halo_properties} using $\alpha_R$ and $q_R$ from \citet{Paranjape_et_al_2018} and $t$ from \citet{Wang_et_al_2011} as proxies for the tidal anisotropy and found that all of these proxies were not as efficient at removing assembly bias as $R_{\rm tidal}$.

\subsection{Measuring gravitational heating}
\label{ssec:grav_heating}

To gauge the contribution of gravitational heating to assembly bias, we use the mass of bound matter within the tidal radius, $R_{\rm tidal}$, defined in the previous section:
\begin{align}
	\label{eq:m_tidal_b}
	M_{\rm tidal,b} = \int_0^{R_{\rm tidal}}dR\int_0^{V_{\rm esc}(R)}dV\frac{dM}{dR\,dV}
\end{align}
Here, $V$ is the absolute velocity of a particle relative to the halo centre, while $V_{\rm esc}(R)$ is the escape velocity at a radius $R$ from the halo centre computed assuming that the halo is well-approximated by an NFW profile:
\begin{align}
	\label{eq:v_esc}
    V_{\rm esc} = V_{\rm vir}\,\left\{2\,\frac{(1 + c_{\rm vir})\,\ln{(1+c_{\rm vir}x)}}{x\,\left[(1 + c_{\rm vir})\,\ln{(1+c_{\rm vir})} - c_{\rm vir}\right]}\right\}^{1/2}.
\end{align}
Here, $x=r/R_{\rm vir},$ $V_{\rm vir}=\sqrt{GM_{\rm vir}/R_{\rm vir}},$ and $c_{\rm vir}$ is halo concentration. To speed up particle containment checks when computing mass profiles, we apply the algorithm described in Appendix \ref{app:fast_halo_containment_checks}.

We also construct the variable
\begin{align}
	\label{eq:m_beta_b}
	M_{\rm \beta,b} = \int_0^{\beta\,R_{\rm vir}}dR\int_0^{V_{\rm esc}(R)}dV\frac{dM}{dR\,dV}
\end{align}
for some constant $\beta.$ $M_{\beta, {\rm b}}$ allows us to isolate the effect of gravitational heating from the effect of external tidal fields because it does not include a dependence on $R_{\rm tidal}.$ Although a range of $\beta$ were used in our analysis, our results are primarily reported in terms of $\beta=3,$ for reasons we describe in section \ref{ssec:env_causes}.

While these approximations are standard practice for computing particle boundedness, it is likely that they break down significantly in the outskirts of haloes. We discuss this in greater depth in Appendix \ref{app:boundedness} and argue that this should not have a significant effect on our results in section \ref{ssec:proxy_def}.

\subsection{Assembly bias statistics}
\label{ssec:assembly_bias_metrics}
To study assembly bias, one must have a statistic that measures how clustering strength depends on a halo age proxy, $c_{\rm vir}$ in our case. The most direct approach is to split haloes into high-$c_{\rm vir}$ and low-$c_{\rm vir}$ samples, measure the clustering strength of each sample independently using correlation functions, and compare them. 
There are multiple ways of doing this, ranging from measuring the two-point correlation function of haloes, $\xi_{\rm hh}$, in each $c_{\rm vir}$-selected subsample to measuring the bias function, $b(r) = \xi_{\rm hm}/\xi_{\rm mm}$ \citep[e.g,][]{Gao_et_al_2005,Gao_White_2007,Faltenbacher_White_2010}.
While this family of approaches is a valid and commonly-used, there are a number of associated issues. First, the definition of subsamples is arbitrary, and the strength of the measured signal depends on this definition somewhat. Second, if small $c_{\rm vir}$ ranges are chosen to maximize signal strength, statistical errors increase due to the comparatively small number of haloes used.

We use an alternative statistic -- the marked correlation function \citep[the MCF,][]{Beisbert_Kerscher_2000,Gottloeber_et_al_2002} -- which avoids this issue and which has been used in a number of assembly bias studies \citep[e.g.,][]{Wechsler_et_al_2006,Villarreal_et_al_2017}. For a sample of objects with assigned mark, $m$, the MCF is computed as:
\begin{equation}
	\mathcal{M}(r) = \frac{\langle m_i m_j \rangle_{i,j\in P(r)} - \langle m\rangle^2}{\langle m^2 \rangle - \langle m \rangle^2}.
\end{equation}
Here, $P(r)$ is the set of all pairs which are separated by a distance within the same radial bin as $r$. Following  \citet{Villarreal_et_al_2017}, we define concentration marks for haloes in narrow circular velocity bins as their percentile within the $c_{\rm vir}$ distribution of that bin. Specifically, we use ten logarithmic bins in $V_{\rm peak}$ from 120 km s$^{-1}$ to 220 km s$^{-1}$. The narrow bin width is required because the $c_{\rm vir}$ distribution is mass-dependent. This, combined with the mass-dependence of clustering, would result in illusory assembly bias signals in any halo sample defined over a sufficiently large mass range. 

\subsection{Measuring the connection between assembly bias and other variables}
\label{ssec:connection}
To evaluate the relative contribution of different physical processes to assembly bias, we need
a way to gauge how strongly proxies for these processes, such as $D_{\rm vir},$ $R_{\rm tidal},$ $M_{\rm \beta,b},$ $M_{\rm tidal},$ or $M_{\rm tidal,b},$ are related to assembly bias. One simple way to do this is to measure the correlation coefficient between $c_{\rm vir}$ and each variable. However, as discussed in section \ref{ssec:correlation},  any approach that relies on measuring the connection between a proxy and formation time has serious issues.

Instead, in this paper, we follow an approach similar to that of \citet{Villarreal_et_al_2017}. We determine the strength of the connection between assembly bias and a proxy $X$ by finding the percentage of haloes ranked by $X$ that need to be removed from the sample to eliminate the assembly bias signal. For example, if 30\% of haloes must be removed according to $X$ before the assembly bias signal is eliminated, but only 5\% of haloes must be removed to achieve this for another proxy, $Y,$ we  conclude that the physical process traced by $Y$ has a more significant contribution to assembly bias than the process traced by $X$.

Specifically, we first sort distinct haloes according to a proxy $X$, then remove a fraction of haloes $f = N_{\rm removed}/N_{\rm tot}$ for a series of $f$ values ranging from $0.01$ to $N_{\rm distinct}/N_{\rm tot}$ in steps of $0.01$. We then define $f_{\rm removed}$ as the minimum $f$ for which the MCF is within $1-\sigma$ of zero. The sample variance of the MCF is estimated by dividing the simulation box into eight octants, computing the MCF in each octant and finding the standard error on these MCFs at a constant $f_{\rm removed}$. Note that $f_{\rm removed}$ is normalized by the total number of haloes and not by the number of distinct haloes to make it easier to combine with different subhalo classification schemes.

We use a similar method to estimate the sample variance of $f_{\rm removed}$ itself, computing $f_{\rm removed}$ for each octant independently and finding the standard error on these values. Note that these errors on $f_{\rm removed}$ account for contributions from sample variance computed using the same octants, which means that while the uncertainties accurately estimate the scatter on measurements in independent boxes, there is likely covariance between the $f_{\rm removed}$ errors measured for different proxies within the same simulation. This means that the uncertainty on the \emph{relative ordering} of $f_{\rm removed}$ values for multiple proxies within a single simulation is likely to be smaller than these errors would estimate. We discuss this further in section \ref{ssec:env_causes}.

When calculating $f_{\rm removed},$ we compute the MCF in the radial range $[4, 8]$ comoving $h^{-1}\,$Mpc. We have repeated all analysis in this paper with several other choices of radial ranges and did not find any significant qualitative difference in results. The primary result of moving to larger radii is that the amplitude of the reference MCF becomes smaller relative to the error, meaning that smaller cutoffs are able to make the signal consistent with zero. Thus, to be conservative, we use a relatively small-radius cutoff. We illustrate this in Fig.~\ref{fig:method_comp_cf}, which shows the MCF after distinct haloes below the $f_{\rm removed}$ cutoff for $M_{\rm tidal,b}/M_{\rm vir}$ have been removed from the sample: the MCF is consistent with zero out to  $18\ h^{-1}$ Mpc. We have repeated all analysis in this paper using several different radial ranges and results remain qualitatively similar.

We note that this method is effective only for assembly bias models in which haloes are initially unbiased or negatively biased but where a small subset of haloes in extreme environments are pushed to older ages by some non-linear process. If, instead, assembly bias is strongly present in all environments, there will be no value of $f_{\rm removed}$ which can remove it. It is known that assembly bias is present across all halo ages \citep[e.g., see fig.~3 in][]{Wechsler_et_al_2006}, so a finding that there are variables with small values of $f_{\rm removed}$ would already put interesting constraints on the physics of assembly bias. We discuss this in more depth in section \ref{ssec:which_haloes}.

\subsubsection{Difference between $f_{\rm removed}$ and age correlation}
\label{ssec:correlation}

A number of previous studies evaluated the contribution of a given physical process with an associated proxy, $X,$
by measuring the correlation between $X$ and a proxy of halo age, $A$ \citep[e.g.,][]{Hahn_et_al_2009,Wang_et_al_2011,Hearin_et_al_2016_2,Salcedo_et_al_2018}. This can be done using the Spearman's rank coefficient, $\rho_S(A,X),$ or by measuring the slope of the average trend $X(A).$
While this approach provides indications of which proxies correlate well with halo age, by itself it cannot be used 
to gauge the relative contribution of different physical processes to assembly bias.
This is because a correlation between age and proxy can only lead to assembly bias if clustering strength {\it also} varies strongly as a function of $X$. Comparison of the proxy--halo age correlation strength thus does not provide enough information to unambiguously gauge the contribution of the corresponding process to assembly bias. For example, $D_{\rm vir}$ and $R_{\rm tidal}/R_{\rm vir}$ have roughly the same level of correlation with $c_{\rm vir},$ but haloes experience wildly different differential clustering with with respect to both variables. Consequently, assembly bias is not connected to these two variables with the same strength.

As an illustration, Table \ref{tab:halo_populations} lists values of $f_{\rm removed}$ and the Spearman's rank correlation coefficient, $\rho_{S},$ between $c_{\rm vir}$ and several different proxies and shows that these two quantities are almost completely unrelated. We therefore strongly recommend against drawing conclusions about assembly bias from measurements of correlation with halo age \citep[see][
for additional discussion and caveats associated with using correlation coefficients in the context of assembly bias]{Mao_et_al_2018}.

\section{Analysis}
\label{sec:analysis}

\subsection{Splashback subhaloes and assembly bias}

\begin{figure*}
    \centering
    \includegraphics[width=\columnwidth]{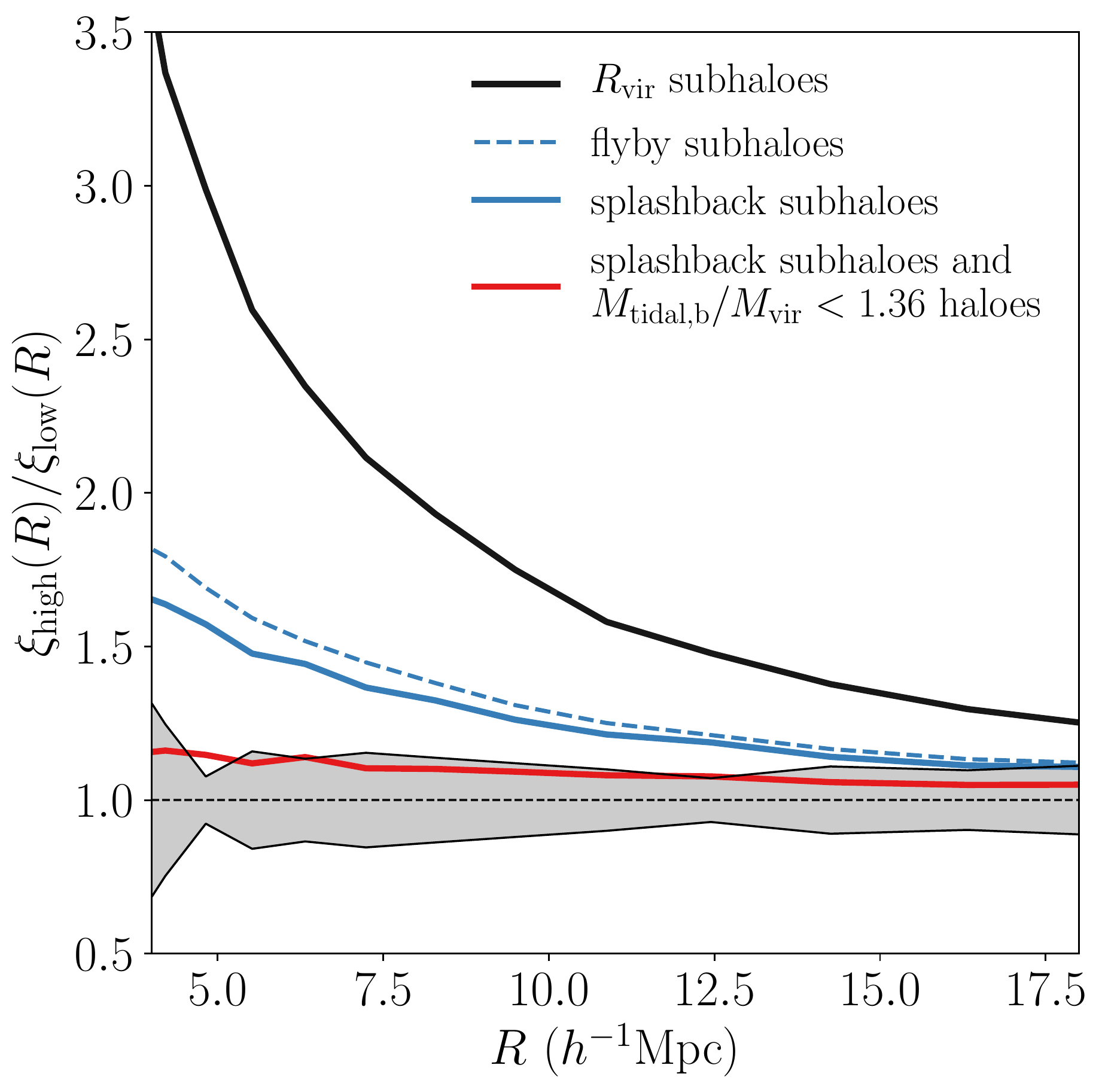}
    \includegraphics[width=\columnwidth]{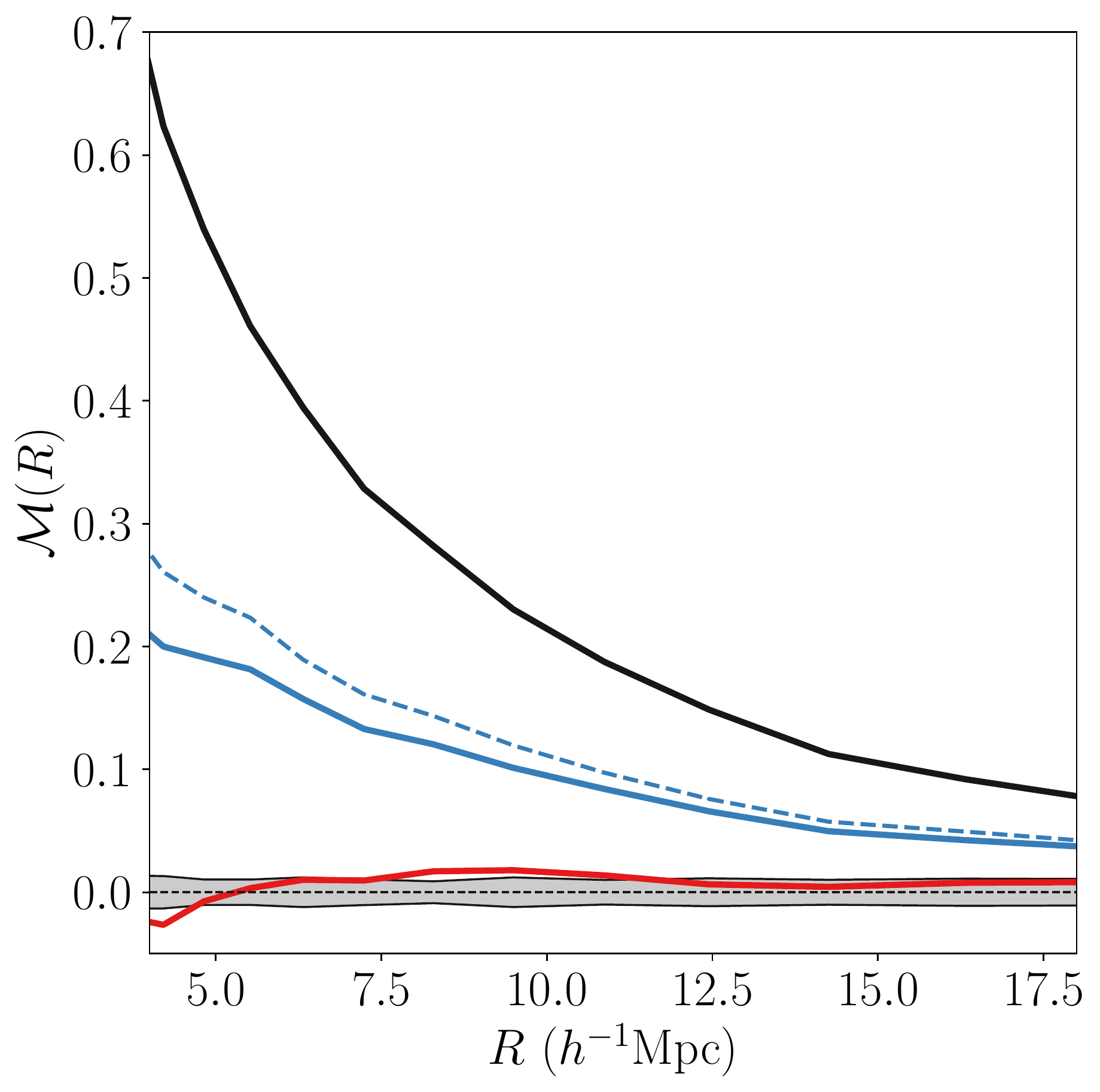}
    \caption
    {		The effect of removing different halo populations on the low-mass
            ($120\ {\rm km\ s}^{-1}<V_{\rm peak}<220\ {\rm km\ s}^{-1}$)
            halo assembly bias signal. The left panel shows assembly bias measured as the ratio
            of the CF of the haloes in the highest and lowest
            15$^{\rm th}$ percentiles of $c_{\rm vir}$, while the right panel shows assembly bias measured as the $c_{\rm vir}$-based MCF.
            Lines are labelled by the groups of haloes which were removed from the sample before
            measurement.
            The grey contours around zero show the 1-$\sigma$ sample variance of the red
            curve. Uncertainties of the three other curves
            are comparable and not shown for visual clarity. Splashback subhaloes
            have been removed in addition to the $M_{\rm tidal,b}$ cut for the red curve.
            Although high-$c_{\rm vir}$ haloes cluster more strongly than
            low-$c_{\rm vir}$ haloes when subhaloes are excluded by
            $R_{\rm vir},$ most of this signal is due to splashback haloes.
            When a small number of tidally truncated haloes (10\% of distinct haloes, 6\% of the total sample)
            are also removed, the difference becomes consistent with zero.
    }
    \label{fig:method_comp_cf}
\end{figure*}

\begin{figure}
   \centering
   \includegraphics[width=\columnwidth]{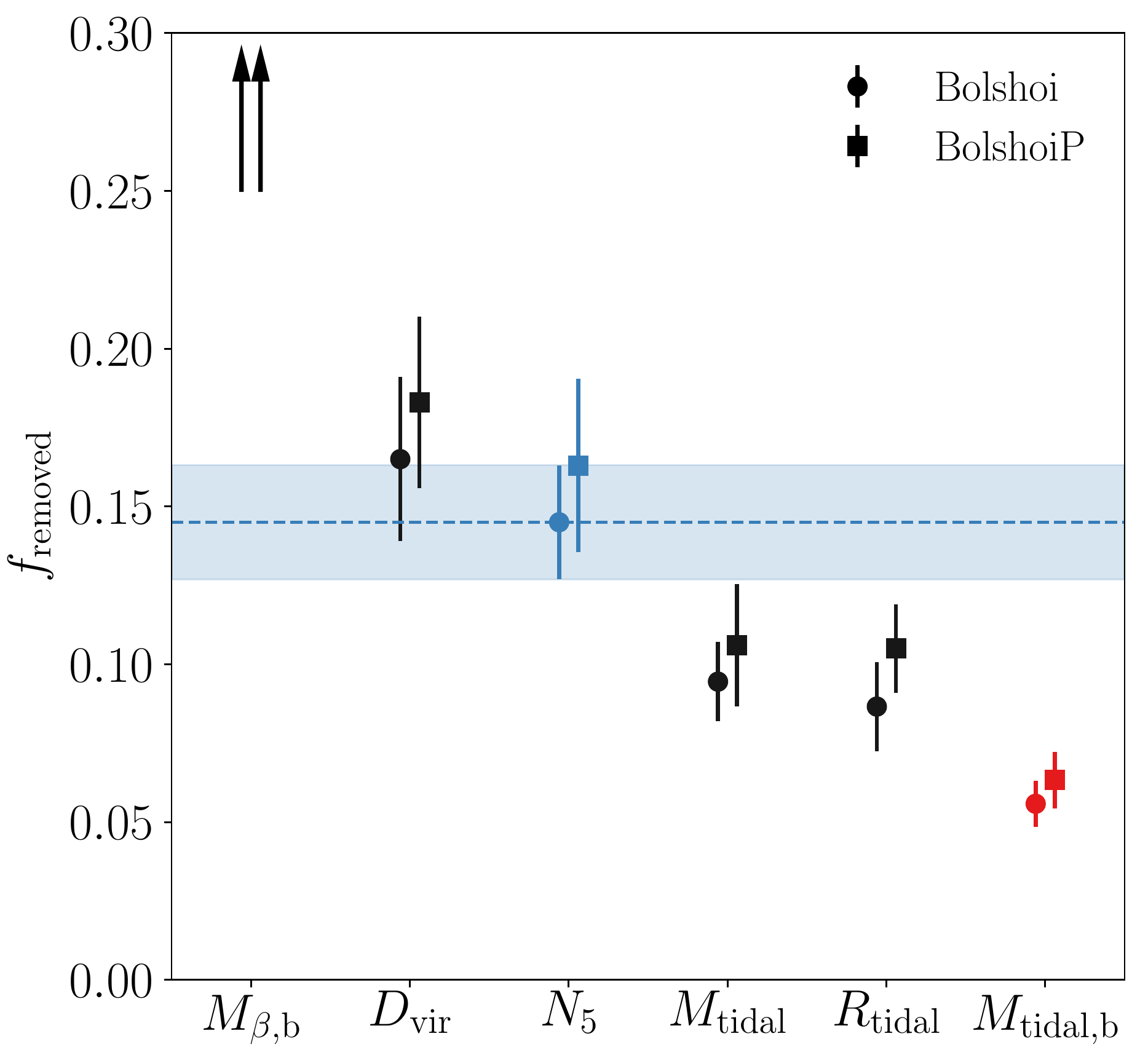}
   \caption{
   The fraction of {\it distinct\/} haloes, $f_{\rm removed}$, rank-ordered by a given physical process proxy that need to be removed to eliminate $c_{\rm vir}$ assembly bias. Note that splashback subhaloes have already been removed from the sample prior to computing $f_{\rm removed}$. The error bars indicate 1-$\sigma$ sample variance. Each quantity listed on the $x$-axis is a proxy for a different physical process: $M_{\beta,b}$ is a proxy for gravitational heating with $\beta$ adjusted to minimize $f_{\rm removed}$ (see \S \ref{ssec:grav_heating}),  $D_{\rm vir}$ is a proxy for single-halo tidal fields (see \S \ref{ssec:singal_halo_tidal_fields}), $N_5$ is an estimate of environmental density (see \S \ref{ssec:env_causes}), $M_{\rm tidal}$ and $R_{\rm tidal}$ are proxies for large-scale tidal fields (see \S \ref{ssec:r_tidal}), and $M_{\rm tidal,b}$ is a proxy for a combination of large scale fields and tidal heating. The $f_{\rm removed}$ values for $M_{\beta,\rm b}$ are outside the vertical range of the plot, which is indicated by arrows (see Tables \ref{tab:halo_populations} and \ref{tab:halo_populations_BolshoiP} for their actual values).  Two proxies have been highlighted with colours: $N_5$ and $M_{\rm tidal,b}$. $N_5$ acts as our control: any variable which has a larger $f_{\rm removed}$ than $N_5$ is more weakly connected to assembly bias than a simple density proxy. A blue band has been added to the figure to make such comparisons easier. $M_{\rm tidal,b}$ is the most effective proxy at eliminating assembly bias, as it requires only $\approx 6\%$ of all haloes (10\% of distinct haloes) to be removed.
   }
   \label{fig:percent_comp}
\end{figure}

We first test whether splashback subhaloes misclassified as distinct haloes by standard subhalo definitions (i.e., splashback subhaloes outside $R_{\rm vir}$ of a larger host) are responsible for low-mass halo assembly bias. The number of haloes removed by our different subhalo definitions is shown in Table \ref{tab:halo_populations}. Our results are shown in Figure~\ref{fig:method_comp_cf}, using both methods discussed in section \ref{ssec:assembly_bias_metrics} for measuring assembly bias. The figure shows that splashback subhaloes cannot account for the entirety of assembly bias, although they contribute about two thirds of the signal. This is consistent with conclusions of the previous studies \citep{Wang_et_al_2009,Sunayama_et_al_2016}. The novel feature of this analysis is that we find a similar effect for two independent definitions of the splashback haloes: using evolutionary trajectories (\S \ref{ssec:dynamical_splashback}) and using non-spherical 3D splashback shells identified using the \textsc{Shellfish} code (\S \ref{ssec:splashback_shells}). 

Note also that although results for the two definitions are similar, the two samples of haloes are not identical. Flyby subhaloes identified using merger trees are guaranteed to have passed their orbital pericentre and thus likely have experienced a strong tidal interaction with the host.
On the other hand, when we use \textsc{Shellfish} all subhaloes within the splashback shell are classified as splashback subhaloes, including those haloes that have entered the shell, but have not yet passed their pericentre. Given that both halo samples have exactly the same local environments, the fact that removal of infalling splashback subhaloes results only in a small decrease of the assembly bias signal means that this portion of the assembly bias signal is due to the stripping subhaloes experience during their pericentre passage. Conversely, any mass growth suppression subhaloes experience on their way to pericentre is comparatively unimportant important. 

We further compare the splashback subhaloes and flyby subhaloes in section \ref{ssec:consitency_of_splashback}.

\subsection{Contribution of tidal truncation and gravitational heating to assembly bias}
\label{ssec:env_causes}
\begin{table}
  \centering
  \caption{The fraction of haloes in the Bolshoi simulation which are removed by the different cuts described in the text. The first three rows show the subhalo fraction, $f_{\rm subhalo},$ for the different subhalo cuts described in section \ref{ssec:subhaloes}. The last six rows correspond to the assembly-bias-removing cuts described in section \ref{ssec:env_causes} for different proxies and show $f_{\rm removed},$ the fraction of haloes which must be removed after splashback subhaloes have been cut from the sample, and $\rho_S,$ the Spearman correlation coefficient between $c_{\rm vir}$ and a given proxy. Note that $f_{\rm removed}$ and $\rho_S(c_{\rm vir}, X)$ are completely uncorrelated, as discussed in section \ref{ssec:correlation}.}
  \label{tab:halo_populations}
  \begin{tabular}{llll}
  \hline\hline
	Subhalo definition & $f_{\rm subhalo}$ &  & Section \\[2mm]
    $R_{\rm vir}$ subhaloes & 0.27 &  & \S \ref{ssec:subhaloes} \\
    flyby subhaloes & 0.33 &  & \S \ref{ssec:dynamical_splashback} \\
    splashback subhaloes & 0.37 &  &\S \ref{ssec:splashback_shells} \\
     \hline
    Removal criterion & $f_{\rm removed}$ & $\rho_S$ & Section\\[2mm]
    $M_{\rm \beta,b}/M_{\rm vir} < 1.24\pm 0.02$ & $0.48\pm 0.04$ & -0.47 & \S\ref{ssec:grav_heating} \\
    $D_{\rm vir} < 4.3\pm 0.4$ & $0.16\pm 0.03$ & -0.16 & \S \ref{ssec:singal_halo_tidal_fields} \\
    $N_5 > 18\pm 2$ & $0.14\pm 0.02$ & 0.09 & \S \ref{ssec:env_causes} \\
	$M_{\rm tidal}/M_{\rm vir} < 1.64\pm 0.04$ & $0.09\pm 0.01$ & -0.23 & \S \ref{ssec:r_tidal} \\
    $R_{\rm tidal}/R_{\rm vir} < 2.8\pm 0.1$ & $0.09\pm 0.01$ & -0.19 & \S \ref{ssec:r_tidal} \\
    $M_{\rm tidal,b}/M_{\rm vir} < 1.36 \pm 0.02$ & $0.056\pm 0.007$ & -0.36 & \S \ref{ssec:grav_heating} \\
  \hline
  \end{tabular}
\end{table}

\begin{table}
  \centering
  \caption{The same as Table \ref{tab:halo_populations}, but for the BolshoiP simulation}
  \label{tab:halo_populations_BolshoiP}
  \begin{tabular}{llll}
  \hline\hline
	Subhalo definition & $f_{\rm subhalo}$ &  & Section \\[2mm]
    $R_{\rm vir}$ subhaloes & 0.28 &  & \S \ref{ssec:subhaloes} \\
    flyby subhaloes & 0.33 &  & \S \ref{ssec:dynamical_splashback} \\
    splashback subhaloes & 0.38 &  &\S \ref{ssec:splashback_shells} \\
     \hline
    Removal criterion & $f_{\rm removed}$ & $\rho_S$ & Section\\[2mm]
    $M_{\rm \beta,b}/M_{\rm vir} < 1.27 \pm 0.08$ & $0.54 \pm 0.07$ & -0.47 & \S\ref{ssec:grav_heating} \\
    $D_{\rm vir} < 4.5 \pm 0.4$ & $0.18 \pm 0.03$ & -0.18 & \S \ref{ssec:singal_halo_tidal_fields} \\
    $N_5 > 21 \pm 2$ & $0.16\pm 0.03$ & 0.09 & \S \ref{ssec:env_causes} \\
    $M_{\rm tidal}/M_{\rm vir} < 1.62 \pm 0.05$ & $0.11\pm 0.02$ & -0.24 & \S \ref{ssec:r_tidal} \\
    $R_{\rm tidal}/R_{\rm vir} < 2.85 \pm 0.1$ & $0.11 \pm 0.01$ & -0.20 & \S \ref{ssec:r_tidal} \\
    $M_{\rm tidal,b}/M_{\rm vir} < 1.36 \pm 0.03$ & $0.062 \pm 0.009$ & -0.36 & \S \ref{ssec:grav_heating} \\
  \hline
  \end{tabular}
\end{table}

We now investigate how the truncation of halo mass growth by the tidal forces, both from a halo's most gravitationally-dominant neighbour and from the entire large-scale matter distribution, contributes to assembly bias. We also investigate the contribution of dynamical heating caused by the collapse of matter into sheets and filaments. To this end we use the five
proxies of these processes defined in sections \ref{ssec:measure_tidal}-\ref{ssec:grav_heating} -- $D_{\rm vir},$ $R_{\rm tidal},$ $M_{\rm \beta,b},$ $M_{\rm tidal},$ and $M_{\rm tidal,b}$ -- and evaluate what fraction of the distinct halo sample ranked by each of the proxies must be removed to eliminate the assembly bias signal.

$D_{\rm vir}$ is the $R_{\rm vir}$-normalized distance to the most tidally dominant halo. It is a proxy of the one-halo contribution to the tidal force proportional to the traditional Hill radius. $R_{\rm tidal}$ is the tidal radius calculated using only the distant matter distribution and $M_{\rm tidal}$ is the mass contained within the tidal radius. $M_{\beta,{\rm b}}$ is the bound mass within $\beta\, R_{\rm vir}$ for a specified constant $\beta$ and serves as a proxy of dynamical heating. Finally, $M_{\rm tidal,b}$ is the bound mass contained within the tidal radius and serves as a proxy for the combined effects of the total tidal force and gravitational heating.

Some care needs to be taken in setting $\beta$ for the proxy $M_{\rm \beta,b}.$ The most straightforward option would be minimize the value of $f_{\rm removed}$ across all values of $\beta,$ but this procedure selects $\beta\approx 1.5,$ which will typically be within the halo's own splashback shell. $M_{\rm \beta,b}$ therefore correlates with $c_{\rm vir}$ simply because the latter determines the mass distribution within the halo. Indeed, we find the Spearman rank coefficient $\rho_S(M(<1.5\times R_{\rm vir}), c_{\rm vir}) = -0.26,$ even before any unbinding procedure has been used. Instead, we choose to set $\beta=3.$ At this distance, correlations between the total enclosed mass and $c_{\rm vir}$ are negligible, and $\beta\, R_{\rm vir}$ will generally be larger than $R_{\rm sp}.$ This choice has little effect on $f_{\rm removed}$, which remains approximately the same for $\beta \gtrsim 2$.

The proxies described above are strongly (anti-)correlated with local matter density. Thus, when we rank-order haloes using these proxies and make cuts, we need to distinguish this procedure from simple density cuts, which do not differentiate between particular physical processes that operate in high-density regions. To this end, we use the number of distinct haloes with 120 ${\rm\,km\,s^{-1}} < V_{\rm peak} < 220 {\rm\,km\,s^{-1}}$ located within $X$ comoving $h^{-1}\,$Mpc of the centre of a halo, $N_X$, as a proxy of the density of the local environment. We tested radii ranging from $1 - 10\,h^{-1}\,{\rm Mpc}$ and found that the assembly bias signal can be eliminated by removing the smallest fraction of haloes for $X=5$. We thus use $N_5$ as our fiducial local environmental density proxy.

In Fig.~\ref{fig:percent_comp} we show the fraction, $f_{\rm removed},$ of all haloes rank-ordered by different proxies that must be removed to eliminate the assembly bias signal (see section \ref{ssec:connection}). The corresponding $f_{\rm removed}$ thresholds for each proxy are presented in Table \ref{tab:halo_populations}, and the red curves in Fig.~\ref{fig:method_comp_cf} show clustering strength as a function of distance after such a cut has been made to $M_{\rm tidal,b}$. Note that statistical errors on the MCF are smaller relative to its amplitude than errors on the $\xi_{\rm high}(r)/\xi_{\rm low}(r)$ curve, which is one of the chief reasons that we use the former in calculations of $f_{\rm removed}$. 

As discussed in section \ref{ssec:connection}, the $1-\sigma$ uncertainties of the MCF shown in Fig.~\ref{fig:percent_comp} are the sample variance uncertainties. Therefore, errors on the \emph{relative ordering} of $f_{\rm removed}$ for multiple proxies within a single box are likely to be smaller than these error estimates. This can be also be seen by comparing the Bolshoi and BolshoiP points in Fig.~\ref{fig:percent_comp}. Although the difference between these two boxes is consistent with the estimated
sample variance, the relative-ordering of the proxies by $f_{\rm removed}$ is quite similar between the boxes, with BolshoiP consistently having $f_{\rm removed}$ values one or two percentage points higher than Bolshoi for all proxies other than the high-scatter $M_{\rm \beta,b}.$ From this we can comfortably infer that the non-systematic error on $f_{\rm removed}$ is on the order of $1\%.$

The first feature apparent in Fig.~\ref{fig:percent_comp} is that it is \emph{possible} to remove assembly bias by making a cut on the local density, meaning that assembly bias is only present in high-density regions. This is consistent with models which predict that low-mass assembly bias is caused by non-linear processes, but is not necessarily a generic prediction of such models, as one could imagine assembly bias existing in all regions to different degrees of severity. Fig.~\ref{fig:percent_comp} also shows that the portion of assembly bias which is not caused by misclassified splashback subhaloes is
due to a small number of haloes in extreme environments: the cut $M_{\rm tidal,b}/M_{\rm vir}<1.36$ removes only 6\% of all haloes but reduces assembly bias to statistically undetectable levels. For comparison, 
the cut to the density proxy $N_5$ removes assembly bias when 14\% of haloes are removed.

Further testing shows that there are two reasons why assembly bias can be eliminated by removing only a small fraction of haloes. First, the mean value of $c_{\rm vir}$ ceases to be a strong function of these proxies once the haloes below the $f_{\rm removed}$ cutoff have been removed from the sample. Second, halo clustering strength varies strongly as a function of proxy value within the cutoff range, but is almost constant throughout the remaining sample.

Finally, Fig.~\ref{fig:percent_comp} shows that $f_{\rm removed}$ for both $D_{\rm vir}$, a proxy for the single-halo tidal force, and $M_{\rm \beta,b}$, a proxy for
dynamical heating, are at least as large as $f_{\rm removed}$ for $N_5$. Even if $M_{\rm \beta,b}$ uses values of $\beta$ small enough that it is primarily picking up features in the halo's own density profile, $f_{\rm removed}$ stays above 0.25. This means that the effect of single-halo tidal forces and dynamical heating on assembly bias cannot be distinguished from the trivial effect of environmental density on halo bias, which means that neither can account for the assembly bias on their own.

In contrast, $f_{\rm removed}$ for $R_{\rm tidal}$, $M_{\rm tidal}$, and $M_{\rm tidal,b}$ are smaller than for $N_5,$ indicating that these proxies are more closely connected to assembly bias than local density. The fact that $R_{\rm tidal}$ and $M_{\rm tidal},$ which are calculated using only the large-scale contribution to the tidal field, have lower $f_{\rm removed}$ than $D_{\rm vir}$ shows that it is the tidal force from large-scale structures, not from individual haloes, that play the dominant role in the assembly bias. $M_{\rm tidal,b}$ has the lowest $f_{\rm removed}$ and is thus the most closely connected to assembly bias of all the proxies we consider.

To summarize, the results of this and previous subsections  show that  $\approx 70\%$ of the low-mass assembly bias signal in $c_{\rm vir}$ is due to splashback subhaloes. The remaining  $\approx 30\%$ of the signal is due to 10\% of distinct haloes (6\% of all haloes) that are affected by a combination of the truncation of their mass growth by large-scale tidal fields and dynamical heating caused by the collapse of sheets and filaments. There are thus three different physical processes that affect halo mass growth which all contribute significantly to the assembly bias signal.

\subsection{The spatial and concentration distributions of the haloes responsible for assembly bias}
\label{ssec:which_haloes}

\begin{figure*}
   \centering
   \includegraphics[width=2\columnwidth]{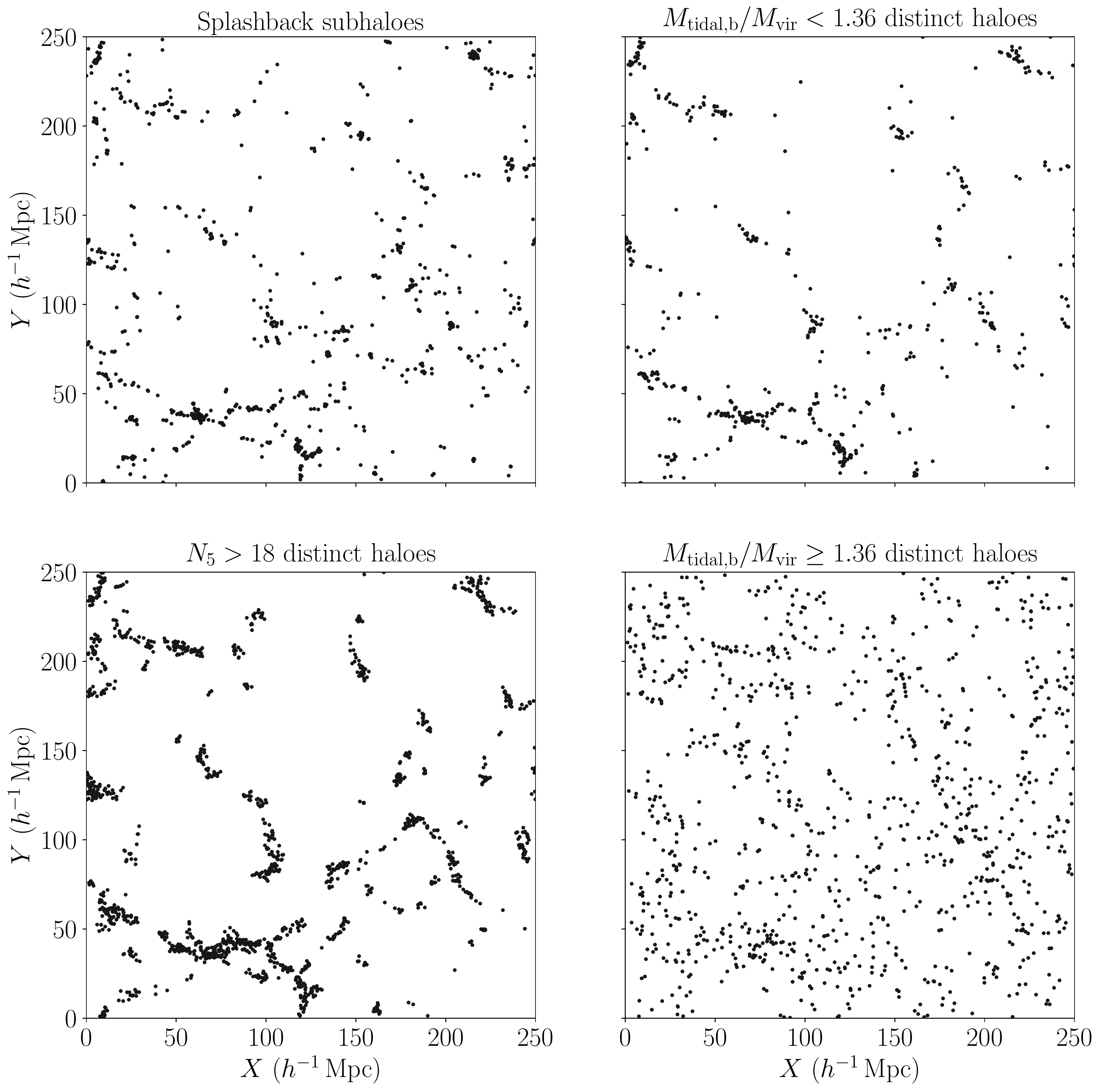}
   \caption{
            The spatial distribution of different classes of haloes within a
            25 $h^{-1}$Mpc thick slice of the Bolshoi simulation.
            The top left panel shows the location of splashback subhaloes outside
            $R_{\rm vir}$ of their hosts, the top right panel shows distinct haloes with $M_{\rm tidal,b}/M_{\rm vir}<1.36,$
            the bottom left panel shows distinct haloes with $N_5 > 18,$ and the bottom right panel shows the remaining haloes after haloes in the two top panels have been removed.
            The haloes in the bottom right panel have been subsampled by a factor of five.
            Note that the assembly bias signal for the haloes in the bottom right panel is consistent with zero.
}
   \label{fig:box_slice}
\end{figure*}

\begin{figure}
   \centering
   \includegraphics[width=\columnwidth]{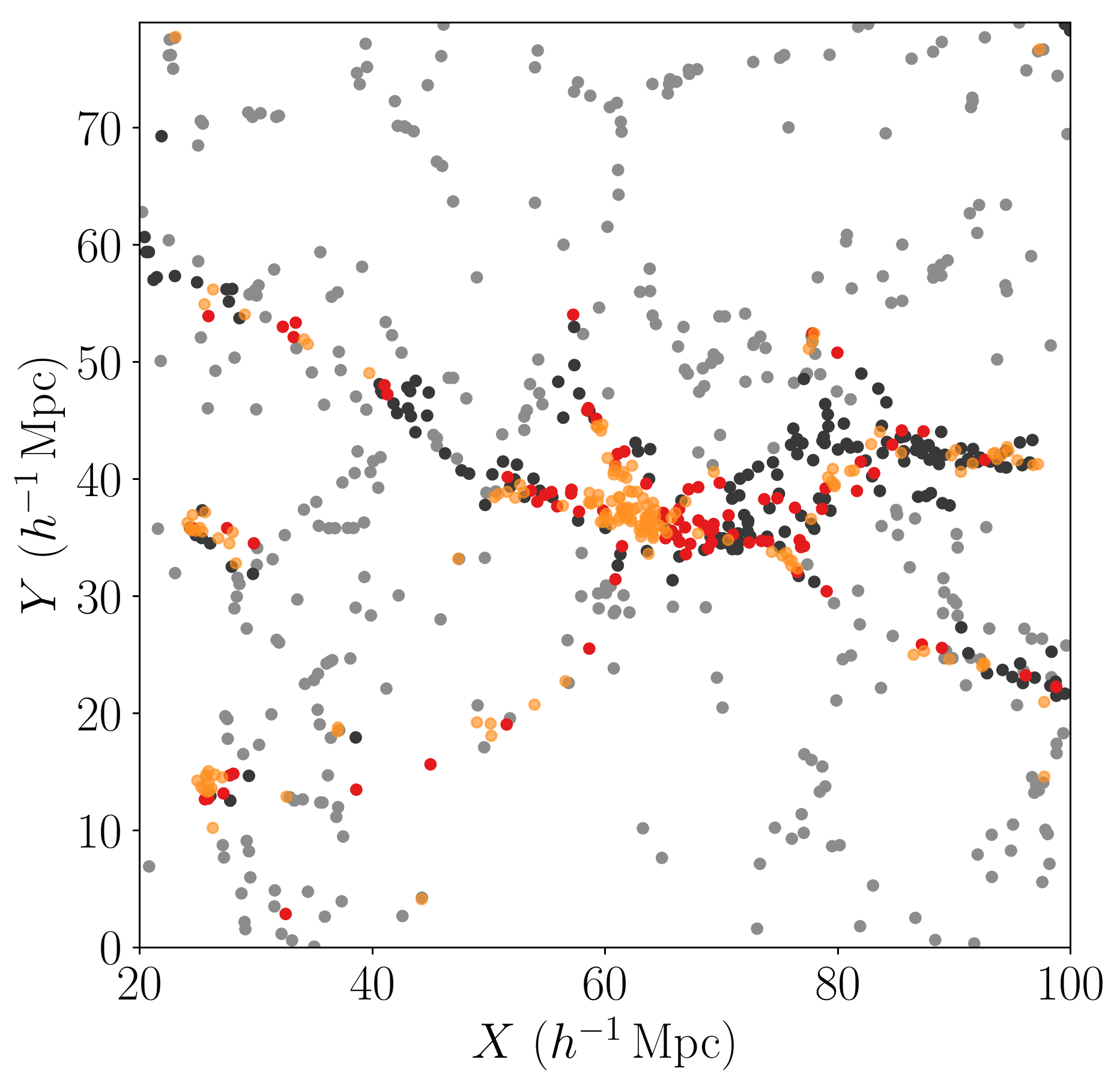}
   \caption{
   		A zoomed-in view of the lower left corner of the panels in Fig.~\ref{fig:box_slice} with the haloes that are removed by different criteria plotted with different colors. Splashback subhaloes outside the virial radii of their hosts are shown as orange points, distinct haloes cut using the $M_{\rm tidal,b}$ criterion are shown as red points, haloes removed by the $N_5$ cut are shown as dark grey points, and haloes surviving all of the cuts are shown by the light grey points. The assembly bias signal is consistent with zero when orange and red points are removed. 
}
   \label{fig:zoom_box_slice}
\end{figure}

\begin{figure}
   \centering
   \includegraphics[width=\columnwidth]{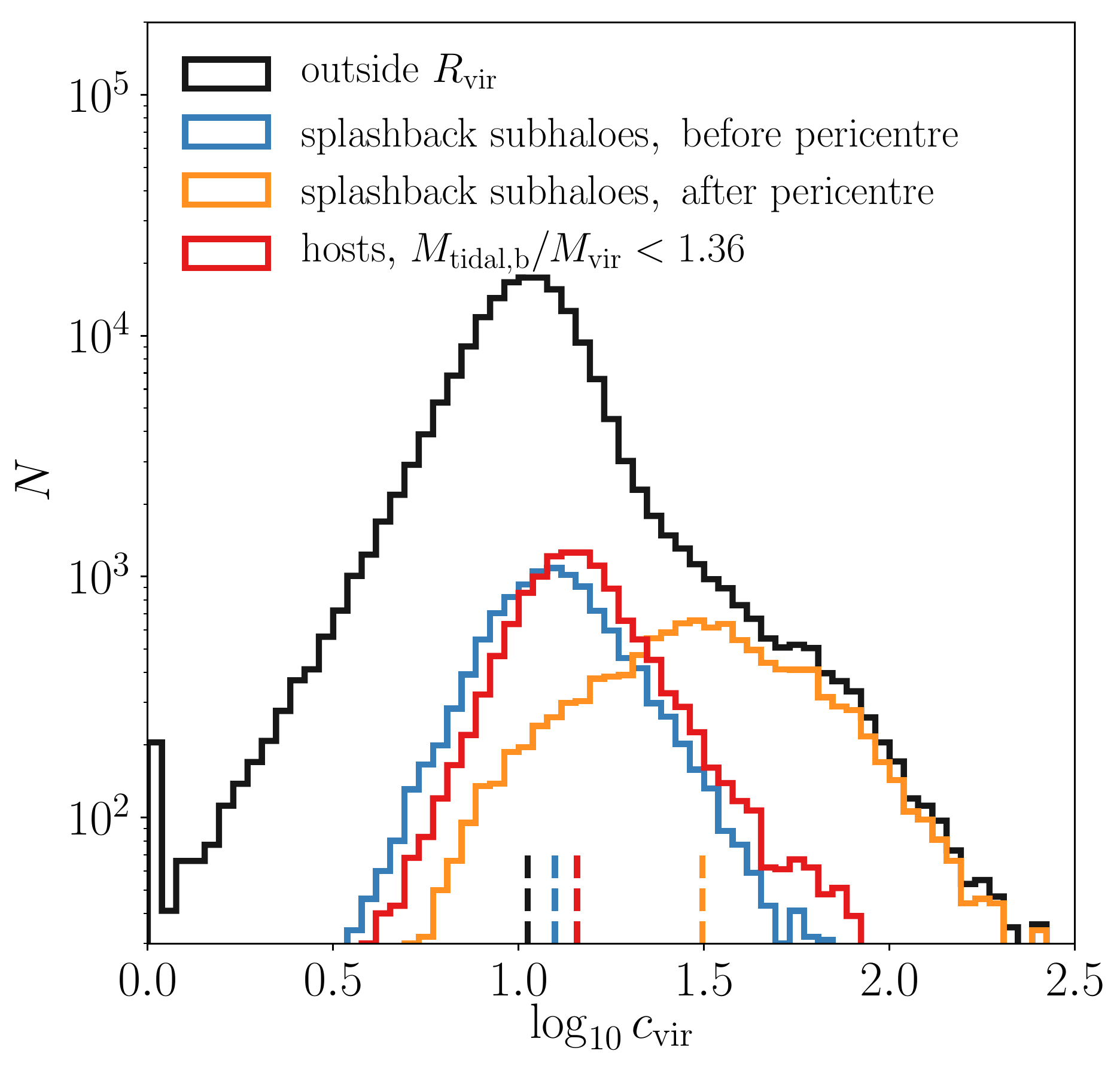}
   \caption{
   The distribution of concentrations for different low-mass halo populations.
   The black curve shows the concentration distribution for all haloes outside
   $R_{\rm vir}$ of any host. 
   The $c_{\rm vir}$ distributions of splashback subhaloes on their first orbit are shown in blue and yellow.
   The blue curve corresponds to haloes on first infall which have passed the splashback shell but not their first pericentre, and the yellow curve
   corresponds to splashback subhaloes which have passed their first pericentre and have re-entered the region between
   $R_{\rm vir}$ and the splashback shell.
   The red curve shows the $c_{\rm vir}$ distribution for haloes with $M_{\rm tidal,b}/M_{\rm vir} < 1.36$.
   Vertical dashed lines show the medians of each distribution.
   Note that when haloes corresponding to the red, yellow, and blue curves have been removed from the general sample, the $c_{\rm vir}$ assembly bias signal is consistent with zero.
  Note also that post-pericentre splashback subhaloes are responsible for almost all of the high-$c_{\rm vir}$ tail of the general population.
   }
   \label{fig:c_distr}
\end{figure}

In Fig.~\ref{fig:box_slice} we show the spatial distribution of splashback subhaloes located outside $R_{\rm vir}$ of their hosts in a 25 $h^{-1}$ Mpc thick slice of the Bolshoi simulation volume. In the same volume, we show the sets of distinct haloes that are removed 
under the criteria $N_5>18$ and $M_{\rm tidal,b}/M_{\rm vir}<1.36$, which each independently eliminate assembly bias. We also show the spatial distribution of a random $\times 5$ subsample of the set of distinct haloes that were not removed by the $M_{\rm tidal,b}/M_{\rm vir}<1.36$ cut. 

Fig.~\ref{fig:box_slice} shows that both splashback subhaloes and haloes with low $M_{\rm tidal,b}/M_{\rm vir}$ values are strongly clustered in the fabric of the cosmic web: they lie within filaments, sheets, and nodes with characteristic scales of tens of $h^{-1}$ Mpc. Splashback subhaloes cluster strongly because they trace the spatial distribution of their massive host haloes, which are predominantly found in these dense environments. Haloes with low $M_{\rm tidal,b}/M_{\rm vir}$, on the other hand, are strongly clustered because the two physical processes that reduce $M_{\rm tidal, b}/M_{\rm vir}$ -- strong tidal forces and gravitational heating -- are strongest in similarly dense regions. The distributions of haloes with larger values of $M_{\rm tidal, b}/M_{\rm vir}$ or smaller values of $N_5$ are less clustered. We also provide a zoomed-in view of the distribution of these different groups in Figure~\ref{fig:zoom_box_slice}.

Fig.~\ref{fig:c_distr} shows concentration distributions for different groups of haloes: all haloes outside $R_{\rm vir}$ of any host, splashback subhaloes outside $R_{\rm vir}$ that have not passed through pericentre of their orbit, splashback subhaloes outside $R_{\rm vir}$ which have passed their first pericentre, and distinct haloes outside the splashback shell of any host which have low $M_{\rm tidal,b}/M_{\rm vir}.$ This figure shows that the $c_{\rm vir}$ distribution of post-pericentre splashback subhaloes is biased to much larger values and are responsible for almost the entire  high-$c_{\rm vir}$ tail of the overall concentration distribution. This indicates that the concentrations of such halos are affected substantially by the strong tidal interaction they experienced during their pericentre passage, which strips mass preferentially at the outskirts of haloes, thereby increasing their concentration \citep[e.g.,][]{Kazantzidis_et_al_2004}. In contrast, splashback subhaloes that are on their first infall and distinct haloes with $M_{\rm tidal,b}/M_{\rm vir}<1.36$ have comparable concentration distributions and are only slightly shifted relative to the overall distribution of concentrations. The modest shift in $c_{\rm vir}$ is  consistent with an older age of these haloes, rather than the large 
concentration boost in halos that have experienced tidal stripping. We note that the strength of the high $c_{\rm vir}$ tail becomes weaker if the halo sample is defined by $M_{\rm vir}$ or $V_{\rm max}$. This is because haloes that lost mass after their first pericentre passage drop out of the $M_{\rm vir}$-defined sample, but stay within the $V_{\rm peak}$ defined sample.

\subsection{Time and mass dependence of assembly bias}
\label{ssec:time_evolution_of_assembly_bias}

\begin{figure}
   \centering
   \includegraphics[width=\columnwidth]{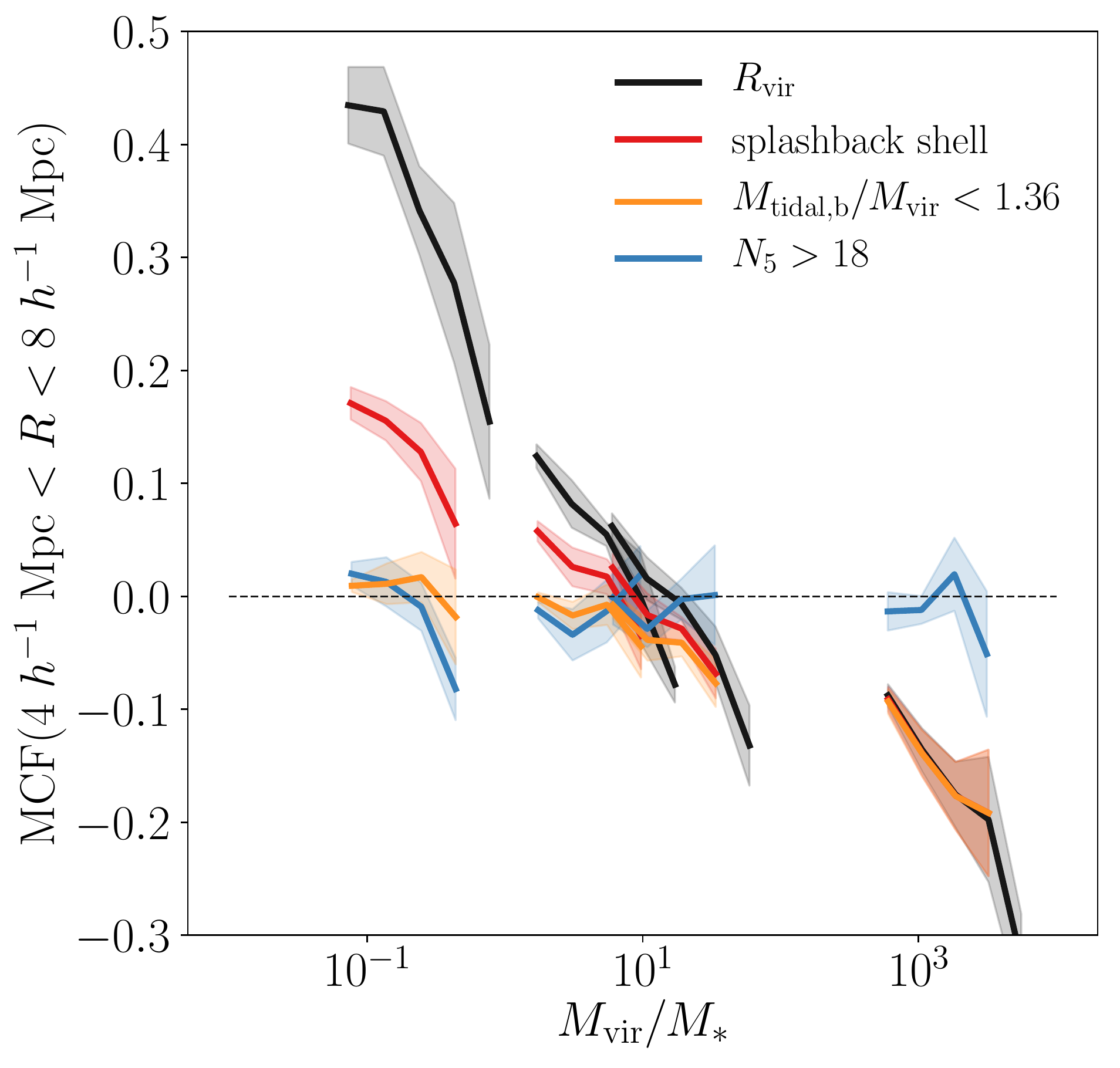}
   \caption{The dependence of assembly bias on $M_{\rm vir},$ scaled by the
            non-linear collapse mass scale, $M_{\rm *}.$ This plot was created
            from the $z$ = 0, 1, 1.4, and 3 snapshots of Bolshoi and shows the median values
            of $M_{\rm vir}/M_*$ in thin $V_{\rm peak}$ bins versus
            the MCF between 4 $h^{-1}$ Mpc and 8 $h^{-1}$ Mpc for
            each bin.
            $1-\sigma$ sample variance  of the MCFs are plotted as shaded regions. Lines of different colour show MCFs for halo samples with different cuts indicated in the legend. $V_{\rm peak}$ bins below our
            convergence limit of 120 km s$^{-1}$ and bins with errors on
            the MCF larger than 0.1 are not plotted (this typically occurs at
            $V_{\rm peak}\approx$ 300 km s$^{-1}$). 
            Non-linear effects strongly reduce assembly bias at
            low masses but have no impact on high-mass assembly bias because this
            effect has a different physical origin. However, a single cut to local density is effective
            at removing assembly bias at all masses.
            }
   \label{fig:z_trend}
\end{figure}

As discussed in section \ref{sec:introduction}, $c_{\rm vir}$ halo assembly bias has opposite signs at masses above and below the non-linear collapse mass scale, $M_*.$ 
Fig.~\ref{fig:z_trend} shows the dependence of assembly bias on $M_{\rm vir}/M_*$ in the Bolshoi simulation for the entire sample of distinct haloes (black line and shading), and samples in which subsets of haloes have been removed using different criteria discussed earlier in this section (coloured lines). 
We first divide haloes into logarithmic bins of  $V_{\rm peak}$ with 0.08 dex width. We use bins of $V_{\rm peak}$ to be consistent with the rest of our analysis, although we show the assembly bias signal as a function of the median $M_{\rm vir}/M_*$ within each bin. For each bin above the convergence limit of $V_{\rm peak}=120\,\rm km\,s^{-1}$ we  measure the MCF in the separation range of $4-8$ comoving $h^{-1}$ Mpc and split the simulation into eight equal-size sub-boxes to estimate the 1-$\sigma$ error on the MCF amplitude. To probe a wide range of $M_{\rm vir}/M_{\rm *}$ values, we use the $z = 0$, $1$, $1.4$, and $3$  Bolshoi snapshots, with the $z=0$ snapshot giving us access to the lowest values of $M_{\rm vir}/M_*$ and $z=3$ giving us access to the highest. 

The dependence of assembly bias on $M_{\rm vir}/M_*$ for distinct haloes outside $R_{\rm vir}$ of any larger host is consistent with the results of \citet{Wechsler_et_al_2006}.
Removing splashback subhaloes reduces the assembly bias substantially at $M_{\rm vir}/M_*\lesssim 10$, but does not eliminate it completely, and does not affect assembly bias at larger masses. Removing further distinct haloes using 
$M_{\rm tidal,b}/M_{\rm vir}<1.36$ cut eliminates assembly bias entirely at $M_{\rm vir}/M_*\lesssim 5$, but likewise does not affect the assembly bias at larger $M_{\rm vir}/M_*$. This illustrates that the physical origin of assembly bias in the high-mass regime is not related to tidal forces or dynamical heating.  

Interestingly, Fig.\ref{fig:z_trend} also shows that removing haloes using environmental density, $N_5,$ \emph{does} remove assembly bias at all $M_{\rm vir}/M_*$. Visual inspection reveals that this is because this cut removes the same spatial regions across time.

Given that halos and large-scale structure evolve with time, we also redid this analysis by removing a constant fraction of distinct haloes ranked by $M_{\rm tidal,b}/M_{\rm vir}$ and by $N_5$ at different redshifts rather than using a fixed cut as in Fig.~\ref{fig:z_trend}. The results of such analysis are almost identical, albeit with slightly higher $f_{\rm removed}.$

\subsection{Sensitivity to splashback subhaloes identification method}
\label{ssec:consitency_of_splashback}

As discussed above, we use two different methods to identify splashback subhaloes: 1) haloes that move within $R_{\rm vir}$ of a larger halo at some point during their evolution and (``flyby subhaloes'') 2) haloes located within the splashback shell identified by the \textsc{Shellfish} algorithm (``splashback subhaloes''). It is clear that the samples of subhaloes identified using these methods cannot be identical: haloes that are on their first approach to a host and are already within the splashback shell but are still outside $R_{\rm vir}$ will  be classified as splashback subhaloes by the second method, but not the first.
Conversely, haloes that previously passed within $R_{\rm vir}$ of the host, but are now outside of the splashback shell identified by \textsc{Shellfish} will be identified by the first method, but not the second. 

We find that $\approx 40\%$ of splashback subhaloes ($\approx 4\%$ of all haloes) are not identified as flyby haloes, but only $\approx 6-8\%$ of flyby subhaloes ($0.4-0.5\%$ of all haloes) are not identified as splashback subhaloes. The latter subhaloes are misidentified largely around host haloes below convergence limit of the \textsc{Shellfish} algorithm:  if we restrict this analysis to host haloes that meet the convergence requirements of $N_{\rm 200m} > 5\times 10^4$  and $\Gamma_{\rm DK14} > 0.5$ \citep[see][for details]{Mansfield_et_al_2017}, we find that only $1-2\%$ of flyby haloes ($\lesssim 0.1\%$ of all haloes) are not identified by the second method. This small fraction indicates 
that the splashback shells identified by \textsc{Shellfish} for well-resolved haloes capture the vast majority of the splashback subhaloes 
identified by the traditional subhalo trajectory method. This also indicates that the fraction of subhaloes ejected by three-body interactions via the slingshot process beyond the splashback shell \citep{Kravtsov_et_al_2004,Sales_et_al_2007} is quite small and that most of the subhaloes outside $R_{\rm vir}$ are on their natural dynamical orbit around their host halo.  
We note that this conclusion should not be extended to $V_{\rm peak} < 120$ without further testing: it is plausible that slingshot processes become more significant at lower masses.

This is consistent with earlier studies that analysed the radial distribution of flyby subhaloes \citep{Ludlow_et_al_2009,Wang_et_al_2009,Li_et_al_2013} and found that flyby subhaloes are common at distances up to  $2\,R_{200c}$, with numbers decreasing quickly at larger radii but with a small population present out to $\approx 4\,R_{\rm 200c}$. We find that the radial distribution of flyby subhaloes is due to large size of the splashback shell relative to $R_{\rm 200c}$, its non-spherical shape, and the substantial scatter between $R_{\rm 200c}$ and the maximum radius of the splashback shell. For our sample of distinct haloes, the mean value of $R_{\rm sp}/R_{\rm 200c}$ is 2.16 and the mean value of $R_{\rm sp,max}/R_{200c}$ is 2.80 with a $1-\sigma$ scatter of $\approx 0.6$, where  $R_{\rm sp, max}$ is the maximum radius of any point on the splashback shell. 

Lastly, Figure~\ref{fig:method_comp_cf} compares the MCF after both methods have been used to remove splashback subhaloes. The difference is small relative to the overall amplitude of the signal. We also find that when the procedure described in sections \ref{ssec:connection} and \ref{ssec:correlation} is used, both classification schemes find similar cutoff values. However, $f_{\rm removed}$ is necessarily $\approx 3\%$ larger when flyby flagging is used to remove splashback subhaloes because these cuts must also remove infalling splashback subhaloes. The exception to this is the $M_{\rm tidal}$ proxy, which is higher for almost all splashback subhaloes than it is for almost all distinct haloes.  Thus, $M_{\rm tidal}$ cannot remove assembly bias without removing the entire sample. This leads us to conclude that our general results are robust to differences in the subhalo classification scheme.

\subsection{Comparison of Bolshoi and BolshoiP simulations}
\label{ssec:bolshoip}

All analysis presented above was done using the Bolshoi simulation with cosmological parameters consistent with the final WMAP mission constraints (see section \ref{ssec:simulations_and_codes}). To test the dependence of our results on the assumed cosmology, we repeated all analysis using the BolshoiP simulation, which assumes cosmological parameters consistent with the Planck mission constraints and found that all of the results are qualitatively consistent.  The difference in $\Omega_{\rm m}$ in the Bolshoi and BolshoiP simulations leads to small changes in the cutoff values for $M_{\rm \beta,b}$, $D_{\rm vir},$ $M_{\rm tidal}$, $R_{\rm tidal},$ and $M_{\rm tidal,b}$, but the values of $f_{\rm removed}$ are within 0.01 of the values found for the  Bolshoi simulation for all cuts, which the exception of the high-error $f_{\rm removed}$ value for our least efficient proxy, $M_{\rm \beta,b}.$

\section{Discussion}
\label{sec:discussion}

\subsection{Issues associated with proxy definitions}
\label{ssec:proxy_def}

In this study we define and use several proxies of physical processes that could conceivably contribute to assembly bias. Of these, $D_{\rm vir},\,R_{\rm tidal},\,M_{\rm tidal},\,M_{\rm\beta,b},$ and $M_{\rm tidal,b}$ require estimating the local tidal force and/or determining whether a given particle is bound or unbound. However, it is not trivial to accurately determine whether a particle is bound in the outskirts of haloes \citep[see, e.g.,][for an extended discussion of related issues]{Behroozi_et_al_2013}, and strong assumptions and approximations must be employed in the estimates of tidal forces. 
Errors made in estimating a particular proxy should result in additional scatter in its correlation with $c_{\rm vir}$ and should increase the uncertainty in our estimate of its contribution to the assembly bias. As a corollary, \emph{improvements in proxy definitions should only decrease the measured $f_{\rm removed}$ values.} In practice, only  $M_{\rm tidal,b}$ is strongly affected both by uncertainties in the tidal force estimate and by issues of identifying bound particles, which means that improvements in proxy estimates would primarily reduce $f_{\rm removed}$ for  $M_{\rm tidal,b}$, while having an equal or lesser effect on our other proxies. This means that such improvements would not change our conclusions.

A detailed analysis of the errors associated with the approximations necessary for tidal force calculation can be found in Appendix \ref{app:tidal_forces}. A discussion of the issues related to
identification of bound particles can be found in Appendix \ref{app:boundedness}.

\subsection{Sensitivity of results to definitional choices}
\label{ssec:definition_dependence}

\begin{figure}
    \centering
        \includegraphics[width=\columnwidth]{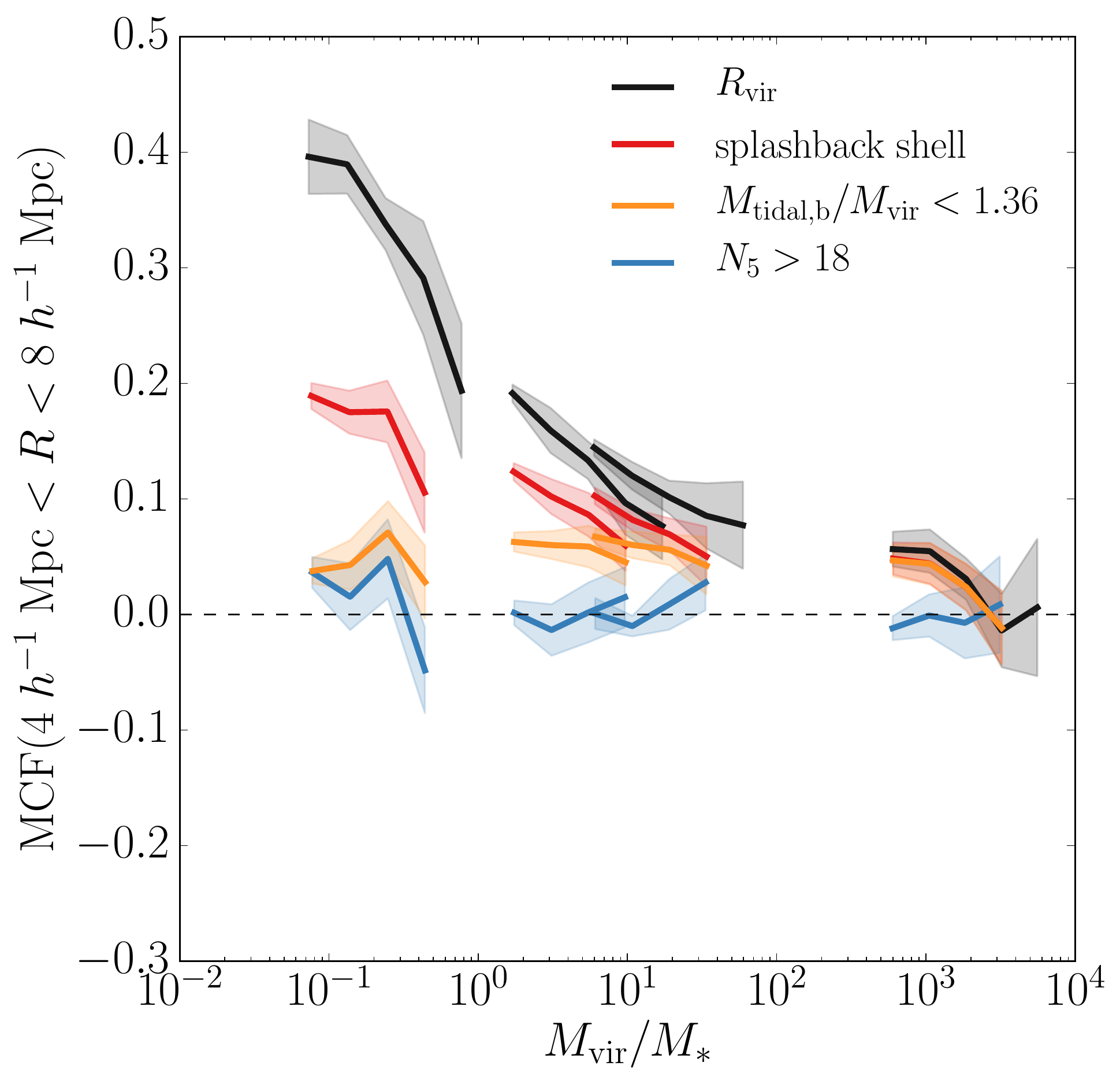}
   \caption{The same as Fig.~\ref{fig:z_trend}, but for MCFs defined in terms of $-a_{1/2}$
   			instead of $c_{\rm vir}$. See section
            \ref{ssec:definition_dependence} for discussion.
            }
   \label{fig:z_trend_a_half}
\end{figure}

In this section we discuss the impact of the choices and assumptions made in our fiducial analysis on our results. We have already discussed how our choice of clustering statistic used to estimate assembly bias affects out results in sections \ref{ssec:assembly_bias_metrics} and \ref{ssec:env_causes} (see Fig.~\ref{fig:method_comp_cf}), so here we focus on the effect of our choices of $V_{\rm peak}$ for defining halo samples, $R_{\rm vir}$ as our reference halo boundary, and halo concentration as our formation time proxy. Although we did present justifications for our choices in section \ref{sec:methods}, it is important to assess how sensitive our results and conclusions are to these choices. 

As an alternative to $V_{\rm peak}$, we could define halo samples using $M_{\rm peak},$ $V_{\rm max}$, or $M_{\rm vir}$. We find that sample selection by $M_{\rm peak}$ leads to results similar to our fiducial case, but using $V_{\rm max}$ or $M_{\rm vir}$ leads to a somewhat different behaviour. The amplitude of the MCF with only $R_{\rm vir}$ subhaloes removed is closer to the amplitude of the MCF with splashback subhaloes removed for a $V_{\rm peak}$ or $M_{\rm peak}$ cut. However, further removing splashback subhaloes with a $V_{\rm max}$ or $M_{\rm vir}$ cut results in only a small decrease in amplitude. This is because subhaloes generally experience significant mass loss and therefore sample
selection based on their peak mass or circular velocity results in larger subhalo fractions compared to selection on current mass \citep[cf., also][]{Nagai_Kravtsov_2005}. The large-$c_{\rm vir}$ tail seen in Fig.~\ref{fig:c_distr} is weaker when $V_{\rm max}$ or $M_{\rm vir}$ is used to define the halo samples for the same reason. Our other results, such as the values of $f_{\rm removed}$ or the spatial distribution of different halo subsets, are largely unaffected. This is because the haloes removed by these cuts have merely had their accretion histories slowed: they have not experienced significant mass loss.

Although most analysis in this paper uses splashback shells as halo boundaries, we use spheres of radius $R_{\rm vir}$ as halo boundaries when we compute fiducial $\mathcal{M}(r)$ curves and when we classify ``flyby'' subhaloes.
We have repeated these analyses using other commonly used values of $\Delta\bar{\rho}_{\rm m}$, and found
that the main difference, unsurprisingly, is in the change of the amplitude of the reference MCF. Definitions with high values of density contrast, such as $\Delta\bar{\rho}_{\rm m}=200\rho_{\rm c}$ or $\Delta\bar{\rho}_{\rm m}=500\rho_{\rm c}$, result in a modest increase of the reference MCF amplitude, while definitions with low density contrasts, such as $\Delta\rho = 200\rho_{\rm m}$, result in a modest decrease of the MCF amplitude.  To decrease the reference MCF amplitude to the level of the MCF after removal of splashback subhaloes requires $\Delta\bar{\rho}_{\rm m} \approx 100\rho_{\rm m}$ -- close to the typical density contrast enclosed by the splashback shell \citep[see Fig. 12 in][]{Mansfield_et_al_2017}. Changes in $\Delta$ used for
radius definition have little effect on the amplitude of the MCF when flyby subhaloes have been removed because most subhaloes have their first pericentres at radii well within all of the commonly-used definitions of halo radius.

We have chosen to use $c_{\rm vir}$ as a proxy of halo age, but assembly bias behaves differently for different proxies \citep[e.g.,][]{Villarreal_et_al_2017,Mao_et_al_2018,Salcedo_et_al_2018}, so one would reasonably wonder if halo removal criteria also depend on this choice. A full investigation of different definitions is beyond the scope of this work, but as a preliminary discussion, we repeat our analysis for the most commonly used alternative age proxy -- the expansion factor at which the virial mass of the main progenitor of a halo was half of the halo's current mass, $a_{1/2}$:
\begin{align}
    M_{\rm vir}(a_{1/2}) &= \frac{1}{2}\,M_{\rm vir}(a_{\rm current}).
\end{align}
Note that although large (small) $c_{\rm vir}$ values correspond to old (young) haloes, the opposite is true for $a_{1/2}.$ To simplify comparison with $c_{\rm vir}$-based results, we use $-a_{1/2}$ as the formation time proxy, so the sign of the MCF retains the same qualitative meaning.

Results for the $-a_{1/2}$ proxy are shown in Fig.~\ref{fig:z_trend_a_half}, where we use the same cuts that removed the assembly bias in the $c_{\rm vir}$-based analysis. The figure shows that stricter cuts are required to eliminate $-a_{1/2}$ assembly bias. When we follow the procedure described in section \ref{ssec:connection}, the $M_{\rm tidal,b}$ cut that removes $-a_{1/2}$ assembly bias results in $f_{\rm removed}=0.14,$ compared to $f_{\rm removed}=0.06$ for $c_{\rm vir}$ assembly bias. Other proxies experience similar increases in $f_{\rm removed},$ with the exception of $N_5,$ which removes assembly bias for both definitions in almost all bins. The fact that Bolshoi and BolshoiP measurements of $f_{\rm removed}$ agree to within 0.01 for all variables (see \S \ref{ssec:bolshoip}) indicates that the differences in $f_{\rm removed}$ between the $c_{\rm vir}$ and $-a_{1/2}$ definitions are significant.

Another difference is that in contrast to the $c_{\rm vir}$ MCF, the amplitude of the $-a_{1/2}$ MCF  does not reverse sign at large halo masses
(see Fig.~\ref{fig:z_trend_a_half} and several previous studies: \citealt{Gao_et_al_2005,Gao_White_2007,Wetzel_et_al_2007,Mao_et_al_2018,Sato-Polito_et_al_2018}).
This puzzling behaviour at first appears to be inconsistent with the physical origin of the high-mass assembly bias argued for by \citet{Dalal_et_al_2008}. However, \citet{Chue_et_al_2018} showed that $a_{1/2}$ and similar measures of formation time can be problematic if measured relative to a standard overdensity mass. Because these definitions do not account for mass in the splashback shell, haloes measured at a constant $M_{\rm vir}$ actually have a range of ``true'' splashback-enclosed masses, and the high-mass haloes will preferentially have early $a_{1/2},$ which increases the level of bias measured for early $a_{1/2}$ haloes. The intersection of this behaviour with Fig.~\ref{fig:z_trend_a_half} requires further study.

\subsection{Comparison with previous work}
\label{ssec:comparison_with_previous_analysis}

The effect of splashback subhaloes on assembly bias was investigated in a number of recent studies \citep{Wang_et_al_2009,Li_et_al_2013,Sunayama_et_al_2016}, which concluded that splashback subhaloes contribute significantly to low-mass halo assembly bias, but cannot account for the entire signal. These studies used the ``flyby'' approach to classify subhalos (similar to the method described in \ref{ssec:dynamical_splashback}), which can include bona fide splashback subhaloes, as well as subhaloes ejected via slingshot effect after dynamical interactions with other subhalos. However, this classification cannot account for a large number of subhaloes within splashback shells that are on their first infall. This left open the question of the contribution of such  infalling subhaloes on assembly bias. In this study we answer this question in section \ref{ssec:consitency_of_splashback}. 

Additionally, these studies have not demonstrated the physical origin of the remaining assembly bias signal. Our results differ from those of \citet{Sunayama_et_al_2016}, which find that splashback subhaloes have little effect on the MCF at large distances ($R\,\gtrsim\,10\,h^{-1}\,$Mpc).  \citet{Sunayama_et_al_2016} used the same simulation and underlying halo catalogues as this paper, so this difference is likely due to the fact that their samples are defined by $M_{\rm vir}$ (see section \ref{ssec:definition_dependence}) and their use of halo bias ratios to measure assembly bias. This statistic results in larger errors than the MCF, as we discussed in \S~\ref{ssec:assembly_bias_metrics} and \ref{ssec:env_causes}. For example, comparison of the bin-to-bin scatter in Fig.~3 and Fig.~4 of \citet{Sunayama_et_al_2016} to the 1-$\sigma$ error contours in the right panel of our Fig.~\ref{fig:method_comp_cf} indicates that their measurements may not have been sensitive enough to probe large-scale assembly bias.

The conjecture that non-linear tidal and dynamical heating effects can be responsible for low-mass halo assembly bias was discussed in a number of studies \citep[e.g.][]{Wang_et_al_2007,Dalal_et_al_2008,Hahn_et_al_2009,Wang_et_al_2011,Hearin_et_al_2016_2,Paranjape_et_al_2018,Salcedo_et_al_2018,Musso_et_al_2018}. Often, arguments for a particular process are based on establishing
existence of a correlation between halo formation time or halo bias and a proxy for a particular process, such as $R_{\rm hill}$, bound mass fraction, the magnitude of tidal eigenvectors, or various measures of tidal anisotropy. While such correlations provide useful information, 
by themselves they are not sufficient to establish that a given physical process is responsible for assembly bias. This is because the proxies of all these processes are all strongly correlated: an explicit comparison, such as that shown in Fig.~\ref{fig:percent_comp}, is more direct and compelling in identifying the responsible process. 

Furthermore, looking at the global connection between halo formation time  and a proxy is problematic for two reasons. First, we show that after splashback subhaloes are removed low-mass halo assembly bias is due to only a small fraction of distinct haloes. Thus, analysis relying on the global correlation strength is not optimal. Second, as discussed in section \ref{ssec:correlation}, the correlation between a proxy and halo formation time by itself contains no information about how closely that proxy is related to assembly bias: a strongly correlated proxy which experiences weak differential clustering, such as $M_{\rm \beta,b}$, will not contribute to assembly bias. We avoid both these issues with the procedure described in section \ref{ssec:connection}.

We find a strong connection between tidal forces from the large-scale mass distribution and assembly bias in agreement with the conclusions of \citet{Hahn_et_al_2009}, \citet{Hearin_et_al_2016_2}, and \citet{Salcedo_et_al_2018}. However, in contrast with these studies, we find that this this effect cannot be effectively approximated by assuming that haloes only feel the tidal force of their most gravitationally dominant neighbor. In fact, we find that when such an approximation is made, the connection is sufficiently weak that it is likely caused simply because the Hill radius is a crude estimate of local density (see section \ref{ssec:env_causes}). This discrepancy is due to two factors. First, some of these studies do not perform the type of multi-variate analysis that would be necessary to differentiate between different contributing physical processes. Second, while these studies effectively map out the the connection between \emph{formation time} and single-halo tidal proxies, this is unrelated to the connection between \emph{assembly bias} and these proxies, an argument we make in section \ref{ssec:correlation}.

Also, contrary to the conclusion of \citet{Paranjape_et_al_2018}, we do not find a compelling evidence that large-scale tidal anisotropy contributes significantly to assembly bias beyond what is expected from its correlation with tidal field strength.
Nevertheless, given the inaccuracies associated with all methods based on second-order approximations to the tidal field (see Appendix \ref{app:tidal_forces}), a more detailed study of tidal field anisotropy could prove fruitful, especially in the context of explicitly studying the tidal environments within structures like sheets and filaments. 

Although our results are in qualitative agreement with the conjectures of \citet{Wang_et_al_2007} and \citet{Dalal_et_al_2008} that gravitational heating is a significant component of assembly bias, we also find that this connection only becomes strong when tracers simultaneously incorporate both gravitational heating and a halo's zone of influence over the local tidal field, such as $M_{\rm tidal,b}.$

Our work uses an approach similar to that of \citet{Villarreal_et_al_2017}, so we have performed an in-depth comparison with their results. We find broad qualitative agreement between our $D_{\rm vir}$ results and the results of \citet{Villarreal_et_al_2017}, but find that quantitatively the $\Delta$ values they report imply $D_{\rm vir}$ values smaller than our findings by $\approx25\%$. 

This difference is due to two factors. First, we find that the sample variance in the boxes used by \citet{Villarreal_et_al_2017} is larger than than they estimated. We estimate the sample variance using subvolumes of the BolshoiP simulation, which has nearly identical mass resolution and cosmology to the  CPla\_L0125 box used by \citet{Villarreal_et_al_2017} and find that the actual variance is larger than the uncertainty they estimated by repeatedly shuffling marks among haloes. This means that MCFs in the CPla\_L0125 box could be lower due to sample variance, which could thus result in less aggressive conditions for the removal of assembly bias.
Second, while the \textsc{Rockstar} halo finder used in \citet{Villarreal_et_al_2017} and in this paper is a state-of-the-art tool for measuring the properties of haloes with density contrasts of $\Delta\gtrsim \Delta_{\rm vir}$ \citep[e.g.,][]{Knebe_et_al_2013}, it cannot effectively measure halo properties at lower density contrasts, such as the $\Delta=20$ contrast used by \citet{Villarreal_et_al_2017}. This is because there is no FOF linking length which can fully percolate all matter out to such large overdensity radii, while also allowing for efficient load-balancing. This leads to underestimates of halo masses and artefacts  in the density profile. Contrary to the findings of \citet{Villarreal_et_al_2017}, we find that even with an unusually large linking length of $b=0.5,$ virtually all haloes have underestimated $M_{\rm 20m}$ masses. The ratio $M_{\rm true}(<R_{\rm 20m,Rockstar})/M_{\rm 20m,all,Rockstar}$ has $1-\sigma$ contours of $1.04-1.13$, with $2\sigma$ fluctuations reaching $\approx 2.5.$ The magnitude of underestimation is significantly worse at more commonly-used linking lengths. This and the fact that haloes no longer follow NFW profiles at large radii \citep[e.g.,][]{Becker_Kravtsov_2011,Diemer_Kravtsov_2014}, adds biases and noise to the $R_{\rm s}$ and $R_{\rm s{\rm Klypin}}$ values measured by \textsc{Rockstar}. This, in turn, artificially reduces the amplitude of the MCF. We discuss this issue in greater depth in Appendix \ref{app:halo_definition_rockstar}.

Our interpretation is consistent with the test presented in Figure 11 of \citet{Villarreal_et_al_2017}, which shows that when $R_{\rm 20m}$ is used to exclude subhaloes, but concentrations are measured from the halo catalogues constructed using larger $\Delta$, the MCF is not consistent with zero. We find that when we replicate their analysis using manually-constructed overdensity profiles, excluding ``subhaloes'' by spheres of radius $R_{\rm 20m}$ is no longer capable of mitigating assembly bias. Larger overdensity radii that are comparable with our reported $D_{\rm vir}$ cutoff values are required.

\subsection{Directions for future work}
\label{ssec:directions_for_future_work}
In this paper, we focus on the dependence of halo bias on $c_{\rm vir}$, but galaxy properties are likely related to a number of halo properties. This means that the effects of secondary biases on galaxy clustering may not be confined to the $c_{\rm vir}$ bias dependence and may remain even if $c_{\rm vir}$ dependence of bias is removed.  As discussed in \ref{ssec:definition_dependence}, this is
true even for an alternative choice of halo formation time proxy, $-a_{1/2}$. Furthermore, \citet{Villarreal_et_al_2017} show that mitigating secondary biases in axis ratio and spin parameter is more difficult than removing bias in $c_{\rm vir}$, although the discussion in section \ref{ssec:comparison_with_previous_analysis} should be kept in mind when assessing these results. It would be useful to perform analysis comparable to the one presented here for a number of other key halo properties to build a more complete understanding of the physical origin of the corresponding dependencies of halo bias. 

One of our key results is that despite the large contribution of tidal forces to $c_{\rm vir}$ assembly bias, this cannot be shown conclusively when using rough and inaccurate estimates of the tidal force, such as $D_{\rm vir}$ or the single-halo $R_{\rm hill}$. Although $R_{\rm tidal}$ defined in section \ref{ssec:r_tidal} accounts for tidal forces from multiple haloes, it still is rather inaccurate, as we show in Appendix \ref{app:tidal_forces} and discuss in \ref{ssec:proxy_def}. The accuracy of the tidal force estimate can be improved by using a higher order approximation of the tidal field, by iteratively recalculating $R_{\rm tidal}$ and removing nearby sources accordingly, or by explicitly evaluating the tidal field outside the halo and identifying the turnover associated with the tidal radius directly. More accurate estimate of the tidal radius could result in a better identification of haloes 
 responsible for assembly bias. This effort could also be aided by incorporating cosmic web classifiers \citep[see review by][]{Libeskind_et_al_2018}, which would allow higher accuracy analytic calculation of the tidal fields associated with nearby large-scale structure, rather than the low-order approximations that are required for generic point distributions.

Finally, all of the tidal field and dynamical heating proxy estimators discussed and used in this paper are computed from simulated quantities and cannot be immediately be applied to observations.  A follow-up exploration of possible observable proxies that can remove particular flavors of secondary halo bias using mock catalogues will be a useful future avenue of research.  

\section{Summary and Conclusions}
\label{sec:summary}

In this study, we present a detailed analysis of the physical causes of assembly bias - the dependence of halo clustering on proxies of halo age. We focus on the origin of halo assembly bias for haloes within the mass range that hosts typical galaxies and on ages traced by $c_{\rm vir}$, but we also present some of our key results across a broad range of masses and for other definitions of age (see Fig.~\ref{fig:z_trend} and \ref{fig:z_trend_a_half}).

We first explore the contribution of ``splashback subhaloes'' to assembly bias, where splashback subhaloes are defined either as halos that have passed within the virial radius of a larger halo at some point in the past or as haloes are located within the splashback shell of a larger halo, as determined by the method of \citet{Mansfield_et_al_2017}. Assembly bias is measured both before and after the removal of these subhaloes.
We show that splashback subhaloes are responsible for about two thirds of the assembly bias signal, but do not account for the entire effect. Moreover, it is the subhaloes that have
passed the pericentre of their orbit at least once that are responsible for the contribution of subhaloes to assembly bias. In addition, we find that the high-$c_{\rm vir}$ tail of the distinct halo distribution is due almost entirely to these same post-pericentre subhaloes.

At the mass ranges considered in this paper, we find that the fraction of haloes which have passed within the splashback shells of their hosts but are later located outside them is small, which indicates that the fraction of haloes ejected beyond the splashback shell due to three-body interactions is small. 
 
We then investigate which additional physical processes contribute to assembly bias. We do this by constructing proxies of these processes for each halo, ranking distinct haloes according to each proxy, and measuring what fraction of the ranked halos need to be removed in order for the assembly bias signal to be statistically consistent with zero. We find that assembly bias is caused by a relatively small number of haloes in dense regions. These haloes have had their accretion histories truncated by a combination of large-scale tidal fields and the high velocities of ambient particles. We also demonstrate that neither process can cause assembly bias on its own and that these tidal fields are not well-modeled by assuming that the dominant tidal contribution comes from a single massive neighbor. We further argue that the commonly-used approach of measuring the correlation between a physical proxy and halo age cannot be used on its own to draw conclusions about the strength of the connection between that proxy and assembly bias.

A key finding of this study is that after splashback subhaloes are removed, the residual assembly bias is due to only 10\% of distinct haloes (5\% of all haloes). To summarize, 27\% of haloes are removed due to a traditional $R_{\rm vir}$-based subhalo cut, a further 10\% are removed due to a splashback subhalo cut, and finally 6\% of all haloes are removed due to the cut based on $M_{\rm tidal,b}/M_{\rm vir}.$ These low $M_{\rm tidal,b}/M_{\rm vir}$ haloes are located within the largest filaments and are only slightly more concentrated than the general population. However, their strong spatial clustering results in an outsized effect on the global assembly bias signal.

We find that in the WMAP cosmology, the removal of haloes above a certain local density, as measured by the
number of haloes within $5\,h^{-1}$ comoving Mpc, $N_5 > 18$, removes assembly bias for both $c_{\rm vir}$ and $a_{1/2}$ across all mass scales and redshifts and that a similar cut exists in a Planck cosmology.
Such a cut removes a much larger fraction of  haloes from the sample than the cut in $M_{\rm tidal,b}/M_{\rm vir}$, and thus does not correspond 
to a real physical process contributing to assembly bias. Nevertheless, this result indicates that it may 
be fruitful to explore whether density-based cuts on mock galaxy catalogs can be used to remove assembly
bias from galaxy samples and motivates further studies in this direction. 

\section*{Acknowledgements}

We would like to thank Antonio Villarreal and Andrew Zentner for useful discussions during this work and Neal Dalal, Andrew Hearin, and Surhud More for comments that improved the paper. We would also like to thank organizers and participants of the "Quantifying and Understanding the Galaxy--Halo Connection" conference held at the Kavli Institute for Theoretical Physics in May 2017 for stimulating discussions and interactions that motivated this work. PM would like to thank Alice Mansfield for moral support while writing this paper. 
AVK would like to thank Institute of Astronomy at Cambridge University and its Sackler visitor program for the warm hospitality during the completion of this paper. The work presented in this paper was supported by the NSF grant AST-1714658 and by the Kavli Institute for Cosmological Physics at the University of Chicago through grant PHY-1125897 and an endowment from the Kavli Foundation and its founder, Fred Kavli. Analyses presented in this paper have been carried out using the Midway cluster at the University of Chicago Research Computing Center, which we acknowledge for support. The CosmoSim database used in this paper is a service by the Leibniz-Institute for Astrophysics Potsdam (AIP). The Bolshoi simulations have been performed within the Bolshoi project of the University of California High-Performance AstroComputing Center (UC-HiPACC) and were run at the NASA Ames Research Center. Analyses presented in this paper were greatly aided by the following free software packages: {\tt NumPy} \citep{numpy_ndarray}, {\tt SciPy} \citep{scipy}, {\tt Matplotlib} \citep{matplotlib}, and \href{https://github.com/}{\tt GitHub}. We have also used the Astrophysics Data Service (\href{http://adsabs.harvard.edu/abstract_service.html}{\tt ADS}) and \href{https://arxiv.org}{\tt arXiv} preprint repository extensively during this project and the writing of the paper.

\bibliographystyle{apj}
\bibliography{ref}

\begin{thebibliography}{95}
\expandafter\ifx\csname natexlab\endcsname\relax\def\natexlab#1{#1}\fi

\bibitem[{{Abazajian} {et~al.}(2005){Abazajian}, {Zheng}, {Zehavi}, {Weinberg},
  {Frieman}, {Berlind}, {Blanton}, {Bahcall}, {Brinkmann}, {Schneider}, \&
  {Tegmark}}]{Abazajian_et_al_2005}
{Abazajian}, K., {et~al.} 2005, \apj, 625, 613

\bibitem[{{Adhikari} {et~al.}(2014){Adhikari}, {Dalal}, \&
  {Chamberlain}}]{Adhikari_et_al_2014}
{Adhikari}, S., {Dalal}, N., \& {Chamberlain}, R.~T. 2014, JCAP, 11, 019

\bibitem[{{Becker} \& {Kravtsov}(2011)}]{Becker_Kravtsov_2011}
{Becker}, M.~R., \& {Kravtsov}, A.~V. 2011, \apj, 740, 25

\bibitem[{{Behroozi} {et~al.}(2013{\natexlab{a}}){Behroozi}, {Loeb}, \&
  {Wechsler}}]{Behroozi_et_al_2013}
{Behroozi}, P.~S., {Loeb}, A., \& {Wechsler}, R.~H. 2013{\natexlab{a}}, \jcap,
  6, 019

\bibitem[{{Behroozi} {et~al.}(2013{\natexlab{b}}){Behroozi}, {Wechsler}, \&
  {Conroy}}]{Behroozi_et_al_2013_2}
{Behroozi}, P.~S., {Wechsler}, R.~H., \& {Conroy}, C. 2013{\natexlab{b}}, \apj,
  770, 57

\bibitem[{{Behroozi} {et~al.}(2014){Behroozi}, {Wechsler}, {Lu}, {Hahn},
  {Busha}, {Klypin}, \& {Primack}}]{Behroozi_et_al_2014}
{Behroozi}, P.~S., {Wechsler}, R.~H., {Lu}, Y., {Hahn}, O., {Busha}, M.~T.,
  {Klypin}, A., \& {Primack}, J.~R. 2014, \apj, 787, 156

\bibitem[{{Behroozi} {et~al.}(2013{\natexlab{c}}){Behroozi}, {Wechsler}, \&
  {Wu}}]{Behroozi_et_al_2013_1}
{Behroozi}, P.~S., {Wechsler}, R.~H., \& {Wu}, H.-Y. 2013{\natexlab{c}}, \apj,
  762, 109

\bibitem[{{Beisbart} \& {Kerscher}(2000)}]{Beisbert_Kerscher_2000}
{Beisbart}, C., \& {Kerscher}, M. 2000, \apj, 545, 6

\bibitem[{{Bertschinger}(1985)}]{Bertschinger_1985}
{Bertschinger}, E. 1985, \apjs, 58, 39

\bibitem[{{Bond} {et~al.}(1996){Bond}, {Kofman}, \&
  {Pogosyan}}]{Bond_et_al_1996}
{Bond}, J.~R., {Kofman}, L., \& {Pogosyan}, D. 1996, \nat, 380, 603

\bibitem[{{Bryan} \& {Norman}(1998)}]{Bryan_Norman_1998}
{Bryan}, G.~L., \& {Norman}, M.~L. 1998, \apj, 495, 80

\bibitem[{{Bullock} {et~al.}(2001){Bullock}, {Kolatt}, {Sigad}, {Somerville},
  {Kravtsov}, {Klypin}, {Primack}, \& {Dekel}}]{Bullock_et_al_2001}
{Bullock}, J.~S., {Kolatt}, T.~S., {Sigad}, Y., {Somerville}, R.~S.,
  {Kravtsov}, A.~V., {Klypin}, A.~A., {Primack}, J.~R., \& {Dekel}, A. 2001,
  \mnras, 321, 559

\bibitem[{{Campbell} {et~al.}(2015){Campbell}, {van den Bosch}, {Hearin},
  {Padmanabhan}, {Berlind}, {Mo}, {Tinker}, \& {Yang}}]{Campbell_et_al_2015}
{Campbell}, D., {van den Bosch}, F.~C., {Hearin}, A., {Padmanabhan}, N.,
  {Berlind}, A., {Mo}, H.~J., {Tinker}, J., \& {Yang}, X. 2015, \mnras, 452,
  444

\bibitem[{{Cautun} {et~al.}(2014){Cautun}, {van de Weygaert}, {Jones}, \&
  {Frenk}}]{Cautun_et_al_2014}
{Cautun}, M., {van de Weygaert}, R., {Jones}, B.~J.~T., \& {Frenk}, C.~S. 2014,
  \mnras, 441, 2923

\bibitem[{{Chue} {et~al.}(2018){Chue}, {Dalal}, \& {White}}]{Chue_et_al_2018}
{Chue}, C.~Y.~R., {Dalal}, N., \& {White}, M. 2018, ArXiv e-prints

\bibitem[{{Dalal} {et~al.}(2010){Dalal}, {Lithwick}, \&
  {Kuhlen}}]{Dalal_et_al_2010}
{Dalal}, N., {Lithwick}, Y., \& {Kuhlen}, M. 2010, ArXiv e-prints

\bibitem[{{Dalal} {et~al.}(2008){Dalal}, {White}, {Bond}, \&
  {Shirokov}}]{Dalal_et_al_2008}
{Dalal}, N., {White}, M., {Bond}, J.~R., \& {Shirokov}, A. 2008, \apj, 687, 12

\bibitem[{{Desjacques}(2008)}]{Desjacques_2008}
{Desjacques}, V. 2008, \mnras, 388, 638

\bibitem[{{Desjacques} {et~al.}(2018){Desjacques}, {Jeong}, \&
  {Schmidt}}]{Desjacques_et_al_2018}
{Desjacques}, V., {Jeong}, D., \& {Schmidt}, F. 2018, \physrep, 733, 1

\bibitem[{{Diemer}(2017{\natexlab{a}})}]{Diemer_2017_2}
{Diemer}, B. 2017{\natexlab{a}}, ArXiv e-prints

\bibitem[{{Diemer}(2017{\natexlab{b}})}]{Diemer_2017_1}
---. 2017{\natexlab{b}}, \apjs, 231, 5

\bibitem[{{Diemer} \& {Joyce}(2018)}]{Diemer_Joyce_2018}
{Diemer}, B., \& {Joyce}, M. 2018, {\mnras} submitted (arXiv/1809.07326)

\bibitem[{{Diemer} \& {Kravtsov}(2014)}]{Diemer_Kravtsov_2014}
{Diemer}, B., \& {Kravtsov}, A.~V. 2014, \apj, 789, 1

\bibitem[{{Diemer} \& {Kravtsov}(2015)}]{Diemer_Kravtsov_2015}
---. 2015, \apj, 799, 108

\bibitem[{{Diemer} {et~al.}(2013{\natexlab{a}}){Diemer}, {Kravtsov}, \&
  {More}}]{Diemer_et_al_2013b}
{Diemer}, B., {Kravtsov}, A.~V., \& {More}, S. 2013{\natexlab{a}}, \apj, 779,
  159

\bibitem[{{Diemer} {et~al.}(2017){Diemer}, {Mansfield}, {Kravtsov}, \&
  {More}}]{Diemer_et_al_2017}
{Diemer}, B., {Mansfield}, P., {Kravtsov}, A.~V., \& {More}, S. 2017, \apj,
  843, 140

\bibitem[{{Diemer} {et~al.}(2013{\natexlab{b}}){Diemer}, {More}, \&
  {Kravtsov}}]{Diemer_et_al_2013}
{Diemer}, B., {More}, S., \& {Kravtsov}, A.~V. 2013{\natexlab{b}}, \apj, 766,
  25

\bibitem[{{Faltenbacher} \& {White}(2010)}]{Faltenbacher_White_2010}
{Faltenbacher}, A., \& {White}, S.~D.~M. 2010, \apj, 708, 469

\bibitem[{{Fillmore} \& {Goldreich}(1984)}]{Fillmore_Goldreich_1984}
{Fillmore}, J.~A., \& {Goldreich}, P. 1984, \apj, 281, 1

\bibitem[{{Gao} {et~al.}(2005){Gao}, {Springel}, \& {White}}]{Gao_et_al_2005}
{Gao}, L., {Springel}, V., \& {White}, S.~D.~M. 2005, \mnras, 363, L66

\bibitem[{{Gao} \& {White}(2007)}]{Gao_White_2007}
{Gao}, L., \& {White}, S.~D.~M. 2007, \mnras, 377, L5

\bibitem[{{Gill} {et~al.}(2005){Gill}, {Knebe}, \& {Gibson}}]{Gill_et_al_2005}
{Gill}, S.~P.~D., {Knebe}, A., \& {Gibson}, B.~K. 2005, \mnras, 356, 1327

\bibitem[{{Gottl{\"o}ber} {et~al.}(2002){Gottl{\"o}ber}, {Kerscher},
  {Kravtsov}, {Faltenbacher}, {Klypin}, \&
  {M{\"u}ller}}]{Gottloeber_et_al_2002}
{Gottl{\"o}ber}, S., {Kerscher}, M., {Kravtsov}, A.~V., {Faltenbacher}, A.,
  {Klypin}, A., \& {M{\"u}ller}, V. 2002, \aap, 387, 778

\bibitem[{{Hahn} {et~al.}(2009){Hahn}, {Porciani}, {Dekel}, \&
  {Carollo}}]{Hahn_et_al_2009}
{Hahn}, O., {Porciani}, C., {Dekel}, A., \& {Carollo}, C.~M. 2009, \mnras, 398,
  1742

\bibitem[{{Han} {et~al.}(2018){Han}, {Li}, {Jing}, {Nishimichi}, {Wang}, \&
  {Jiang}}]{Han_et_al_2018}
{Han}, J., {Li}, Y., {Jing}, Y., {Nishimichi}, T., {Wang}, W., \& {Jiang}, C.
  2018, \mnras

\bibitem[{{Harker} {et~al.}(2006){Harker}, {Cole}, {Helly}, {Frenk}, \&
  {Jenkins}}]{Harker_et_al_2006}
{Harker}, G., {Cole}, S., {Helly}, J., {Frenk}, C., \& {Jenkins}, A. 2006,
  \mnras, 367, 1039

\bibitem[{{Hearin} {et~al.}(2016{\natexlab{a}}){Hearin}, {Tollerud},
  {Robitaille}, {Droettboom}, {Zentner}, {Bray}, {Craig}, {Bradley}, {Barbary},
  {Deil}, {Tan}, {Becker}, {More}, {G{\"u}nther}, \&
  {Sipocz}}]{Hearin_et_al_2016}
{Hearin}, A., {et~al.} 2016{\natexlab{a}}, {Halotools: Galaxy-Halo connection
  models}, Astrophysics Source Code Library

\bibitem[{{Hearin} {et~al.}(2016{\natexlab{b}}){Hearin}, {Behroozi}, \& {van
  den Bosch}}]{Hearin_et_al_2016_2}
{Hearin}, A.~P., {Behroozi}, P.~S., \& {van den Bosch}, F.~C.
  2016{\natexlab{b}}, \mnras, 461, 2135

\bibitem[{{Hearin} {et~al.}(2015){Hearin}, {Watson}, \& {van den
  Bosch}}]{Hearin_et_al_2015}
{Hearin}, A.~P., {Watson}, D.~F., \& {van den Bosch}, F.~C. 2015, \mnras, 452,
  1958

\bibitem[{Hunter(2007)}]{matplotlib}
Hunter, J.~D. 2007, Computing In Science \& Engineering, 9, 90

\bibitem[{{Jing} {et~al.}(2007){Jing}, {Suto}, \& {Mo}}]{Jing_et_al_2007}
{Jing}, Y.~P., {Suto}, Y., \& {Mo}, H.~J. 2007, \apj, 657, 664

\bibitem[{Jones {et~al.}(2001-2016)Jones, Oliphant, Peterson, {et~al.}}]{scipy}
Jones, E., Oliphant, T., Peterson, P., {et~al.} 2001-2016, http://www.scipy.org

\bibitem[{{Kaiser}(1984)}]{Kaiser_1984}
{Kaiser}, N. 1984, \apjl, 284, L9

\bibitem[{{Kazantzidis} {et~al.}(2004){Kazantzidis}, {Mayer}, {Mastropietro},
  {Diemand}, {Stadel}, \& {Moore}}]{Kazantzidis_et_al_2004}
{Kazantzidis}, S., {Mayer}, L., {Mastropietro}, C., {Diemand}, J., {Stadel},
  J., \& {Moore}, B. 2004, \apj, 608, 663

\bibitem[{{Klypin} {et~al.}(2017){Klypin}, {Prada}, \&
  {Comparat}}]{Klypin_et_al_2017}
{Klypin}, A., {Prada}, F., \& {Comparat}, J. 2017, ArXiv e-prints

\bibitem[{{Klypin} {et~al.}(2016){Klypin}, {Yepes}, {Gottl{\"o}ber}, {Prada},
  \& {He{\ss}}}]{Klypin_et_al_2016}
{Klypin}, A., {Yepes}, G., {Gottl{\"o}ber}, S., {Prada}, F., \& {He{\ss}}, S.
  2016, \mnras, 457, 4340

\bibitem[{{Klypin} {et~al.}(2011){Klypin}, {Trujillo-Gomez}, \&
  {Primack}}]{Klypin_et_al_2011}
{Klypin}, A.~A., {Trujillo-Gomez}, S., \& {Primack}, J. 2011, \apj, 740, 102

\bibitem[{{Knebe} {et~al.}(2013){Knebe}, {Pearce}, {Lux}, {Ascasibar},
  {Behroozi}, {Casado}, {Moran}, {Diemand}, {Dolag}, {Dominguez-Tenreiro},
  {Elahi}, {Falck}, {Gottl{\"o}ber}, {Han}, {Klypin}, {Luki{\'c}},
  {Maciejewski}, {McBride}, {Merch{\'a}n}, {Muldrew}, {Neyrinck}, {Onions},
  {Planelles}, {Potter}, {Quilis}, {Rasera}, {Ricker}, {Roy}, {Ruiz},
  {Sgr{\'o}}, {Springel}, {Stadel}, {Sutter}, {Tweed}, \&
  {Zemp}}]{Knebe_et_al_2013}
{Knebe}, A., {et~al.} 2013, \mnras, 435, 1618

\bibitem[{{Kravtsov} {et~al.}(2004){Kravtsov}, {Berlind}, {Wechsler}, {Klypin},
  {Gottl{\"o}ber}, {Allgood}, \& {Primack}}]{Kravtsov_et_al_2004}
{Kravtsov}, A.~V., {Berlind}, A.~A., {Wechsler}, R.~H., {Klypin}, A.~A.,
  {Gottl{\"o}ber}, S., {Allgood}, B., \& {Primack}, J.~R. 2004, \apj, 609, 35

\bibitem[{{Lacey} \& {Cole}(1993)}]{Lacey_Cole_1993}
{Lacey}, C., \& {Cole}, S. 1993, \mnras, 262, 627

\bibitem[{{Li} {et~al.}(2013){Li}, {Gao}, {Xie}, \& {Guo}}]{Li_et_al_2013}
{Li}, R., {Gao}, L., {Xie}, L., \& {Guo}, Q. 2013, \mnras, 435, 3592

\bibitem[{{Li} {et~al.}(2008){Li}, {Mo}, \& {Gao}}]{Li_et_al_2008}
{Li}, Y., {Mo}, H.~J., \& {Gao}, L. 2008, \mnras, 389, 1419

\bibitem[{{Libeskind} {et~al.}(2018){Libeskind}, {van de Weygaert}, {Cautun},
  {Falck}, {Tempel}, {Abel}, {Alpaslan}, {Arag{\'o}n-Calvo}, {Forero-Romero},
  {Gonzalez}, {Gottl{\"o}ber}, {Hahn}, {Hellwing}, {Hoffman}, {Jones},
  {Kitaura}, {Knebe}, {Manti}, {Neyrinck}, {Nuza}, {Padilla}, {Platen},
  {Ramachandra}, {Robotham}, {Saar}, {Shandarin}, {Steinmetz}, {Stoica},
  {Sousbie}, \& {Yepes}}]{Libeskind_et_al_2018}
{Libeskind}, N.~I., {et~al.} 2018, \mnras, 473, 1195

\bibitem[{{Lin} {et~al.}(2016){Lin}, {Mandelbaum}, {Huang}, {Huang}, {Dalal},
  {Diemer}, {Jian}, \& {Kravtsov}}]{Lin_et_al_2016}
{Lin}, Y.-T., {Mandelbaum}, R., {Huang}, Y.-H., {Huang}, H.-J., {Dalal}, N.,
  {Diemer}, B., {Jian}, H.-Y., \& {Kravtsov}, A. 2016, \apj, 819, 119

\bibitem[{{Lu} {et~al.}(2006){Lu}, {Mo}, {Katz}, \& {Weinberg}}]{Lu_et_al_2006}
{Lu}, Y., {Mo}, H.~J., {Katz}, N., \& {Weinberg}, M.~D. 2006, \mnras, 368, 1931

\bibitem[{{Ludlow} {et~al.}(2014){Ludlow}, {Navarro}, {Angulo},
  {Boylan-Kolchin}, {Springel}, {Frenk}, \& {White}}]{Ludlow_et_al_2014}
{Ludlow}, A.~D., {Navarro}, J.~F., {Angulo}, R.~E., {Boylan-Kolchin}, M.,
  {Springel}, V., {Frenk}, C., \& {White}, S.~D.~M. 2014, \mnras, 441, 378

\bibitem[{{Ludlow} {et~al.}(2012){Ludlow}, {Navarro}, {Li}, {Angulo},
  {Boylan-Kolchin}, \& {Bett}}]{Ludlow_et_al_2012}
{Ludlow}, A.~D., {Navarro}, J.~F., {Li}, M., {Angulo}, R.~E., {Boylan-Kolchin},
  M., \& {Bett}, P.~E. 2012, \mnras, 427, 1322

\bibitem[{{Ludlow} {et~al.}(2009){Ludlow}, {Navarro}, {Springel}, {Jenkins},
  {Frenk}, \& {Helmi}}]{Ludlow_et_al_2009}
{Ludlow}, A.~D., {Navarro}, J.~F., {Springel}, V., {Jenkins}, A., {Frenk},
  C.~S., \& {Helmi}, A. 2009, \apj, 692, 931

\bibitem[{{Ludlow} {et~al.}(2018){Ludlow}, {Schaye}, \&
  {Bower}}]{Ludlow_et_al_2018}
{Ludlow}, A.~D., {Schaye}, J., \& {Bower}, R. 2018, arXiv e-prints

\bibitem[{{Ludlow} {et~al.}(2013){Ludlow}, {Navarro}, {Boylan-Kolchin}, {Bett},
  {Angulo}, {Li}, {White}, {Frenk}, \& {Springel}}]{Ludlow_et_al_2013}
{Ludlow}, A.~D., {et~al.} 2013, \mnras, 432, 1103

\bibitem[{{Mansfield} {et~al.}(2017){Mansfield}, {Kravtsov}, \&
  {Diemer}}]{Mansfield_et_al_2017}
{Mansfield}, P., {Kravtsov}, A.~V., \& {Diemer}, B. 2017, \apj, 841, 34

\bibitem[{{Mao} {et~al.}(2018){Mao}, {Zentner}, \& {Wechsler}}]{Mao_et_al_2018}
{Mao}, Y.-Y., {Zentner}, A.~R., \& {Wechsler}, R.~H. 2018, \mnras, 474, 5143

\bibitem[{{Mo} \& {White}(1996)}]{Mo_White_1996}
{Mo}, H.~J., \& {White}, S.~D.~M. 1996, \mnras, 282, 347

\bibitem[{{More} {et~al.}(2015){More}, {Diemer}, \&
  {Kravtsov}}]{More_et_al_2015}
{More}, S., {Diemer}, B., \& {Kravtsov}, A.~V. 2015, \apj, 810, 36

\bibitem[{{More} {et~al.}(2011){More}, {Kravtsov}, {Dalal}, \&
  {Gottl{\"o}ber}}]{More_et_al_2011}
{More}, S., {Kravtsov}, A.~V., {Dalal}, N., \& {Gottl{\"o}ber}, S. 2011, \apjs,
  195, 4

\bibitem[{{Musso} {et~al.}(2018){Musso}, {Cadiou}, {Pichon}, {Codis},
  {Kraljic}, \& {Dubois}}]{Musso_et_al_2018}
{Musso}, M., {Cadiou}, C., {Pichon}, C., {Codis}, S., {Kraljic}, K., \&
  {Dubois}, Y. 2018, \mnras, 476, 4877

\bibitem[{{Nagai} \& {Kravtsov}(2005)}]{Nagai_Kravtsov_2005}
{Nagai}, D., \& {Kravtsov}, A.~V. 2005, \apj, 618, 557

\bibitem[{{Navarro} {et~al.}(1997){Navarro}, {Frenk}, \&
  {White}}]{Navarro_et_al_1997}
{Navarro}, J.~F., {Frenk}, C.~S., \& {White}, S.~D.~M. 1997, \apj, 490, 493

\bibitem[{{Navarro} {et~al.}(2004){Navarro}, {Hayashi}, {Power}, {Jenkins},
  {Frenk}, {White}, {Springel}, {Stadel}, \& {Quinn}}]{Navarro_et_al_2004}
{Navarro}, J.~F., {et~al.} 2004, \mnras, 349, 1039

\bibitem[{{Paranjape} {et~al.}(2018){Paranjape}, {Hahn}, \&
  {Sheth}}]{Paranjape_et_al_2018}
{Paranjape}, A., {Hahn}, O., \& {Sheth}, R.~K. 2018, \mnras, 476, 3631

\bibitem[{{Power} {et~al.}(2003){Power}, {Navarro}, {Jenkins}, {Frenk},
  {White}, {Springel}, {Stadel}, \& {Quinn}}]{Power_et_al_2003}
{Power}, C., {Navarro}, J.~F., {Jenkins}, A., {Frenk}, C.~S., {White},
  S.~D.~M., {Springel}, V., {Stadel}, J., \& {Quinn}, T. 2003, \mnras, 338, 14

\bibitem[{{Reddick} {et~al.}(2013){Reddick}, {Wechsler}, {Tinker}, \&
  {Behroozi}}]{Reddick_et_al_2013}
{Reddick}, R.~M., {Wechsler}, R.~H., {Tinker}, J.~L., \& {Behroozi}, P.~S.
  2013, \apj, 771, 30

\bibitem[{{Salcedo} {et~al.}(2018){Salcedo}, {Maller}, {Berlind}, {Sinha},
  {McBride}, {Behroozi}, {Wechsler}, \& {Weinberg}}]{Salcedo_et_al_2018}
{Salcedo}, A.~N., {Maller}, A.~H., {Berlind}, A.~A., {Sinha}, M., {McBride},
  C.~K., {Behroozi}, P.~S., {Wechsler}, R.~H., \& {Weinberg}, D.~H. 2018,
  \mnras, 475, 4411

\bibitem[{{Sales} {et~al.}(2007){Sales}, {Navarro}, {Abadi}, \&
  {Steinmetz}}]{Sales_et_al_2007}
{Sales}, L.~V., {Navarro}, J.~F., {Abadi}, M.~G., \& {Steinmetz}, M. 2007,
  \mnras, 379, 1475

\bibitem[{{Sandvik} {et~al.}(2007){Sandvik}, {M{\"o}ller}, {Lee}, \&
  {White}}]{Sandvik_et_al_2007}
{Sandvik}, H.~B., {M{\"o}ller}, O., {Lee}, J., \& {White}, S.~D.~M. 2007,
  \mnras, 377, 234

\bibitem[{{Sato-Polito} {et~al.}(2018){Sato-Polito}, {Montero-Dorta}, {Abramo},
  {Prada}, \& {Klypin}}]{Sato-Polito_et_al_2018}
{Sato-Polito}, G., {Montero-Dorta}, A.~D., {Abramo}, L.~R., {Prada}, F., \&
  {Klypin}, A. 2018, ArXiv e-prints

\bibitem[{{Sheth} \& {Tormen}(1999)}]{Sheth_Tormen_1999}
{Sheth}, R.~K., \& {Tormen}, G. 1999, \mnras, 308, 119

\bibitem[{{Springel} {et~al.}(2008){Springel}, {Wang}, {Vogelsberger},
  {Ludlow}, {Jenkins}, {Helmi}, {Navarro}, {Frenk}, \&
  {White}}]{Springel_et_al_2008}
{Springel}, V., {et~al.} 2008, \mnras, 391, 1685

\bibitem[{{Sunayama} {et~al.}(2016){Sunayama}, {Hearin}, {Padmanabhan}, \&
  {Leauthaud}}]{Sunayama_et_al_2016}
{Sunayama}, T., {Hearin}, A.~P., {Padmanabhan}, N., \& {Leauthaud}, A. 2016,
  \mnras, 458, 1510

\bibitem[{{van den Bosch} {et~al.}(2018){van den Bosch}, {Ogiya}, {Hahn}, \&
  {Burkert}}]{van_den_Bosch_et_al_2018}
{van den Bosch}, F.~C., {Ogiya}, G., {Hahn}, O., \& {Burkert}, A. 2018, \mnras,
  474, 3043

\bibitem[{{Van Der Walt} {et~al.}(2011){Van Der Walt}, {Colbert}, \&
  {Varoquaux}}]{numpy_ndarray}
{Van Der Walt}, S., {Colbert}, S.~C., \& {Varoquaux}, G. 2011, ArXiv:1102.1523

\bibitem[{{Villarreal} {et~al.}(2017){Villarreal}, {Zentner}, {Mao}, {Purcell},
  {van den Bosch}, {Diemer}, {Lange}, {Wang}, \&
  {Campbell}}]{Villarreal_et_al_2017}
{Villarreal}, A.~S., {et~al.} 2017, \mnras, 472, 1088

\bibitem[{{Wang} {et~al.}(2009){Wang}, {Mo}, \& {Jing}}]{Wang_et_al_2009}
{Wang}, H., {Mo}, H.~J., \& {Jing}, Y.~P. 2009, \mnras, 396, 2249

\bibitem[{{Wang} {et~al.}(2011){Wang}, {Mo}, {Jing}, {Yang}, \&
  {Wang}}]{Wang_et_al_2011}
{Wang}, H., {Mo}, H.~J., {Jing}, Y.~P., {Yang}, X., \& {Wang}, Y. 2011, \mnras,
  413, 1973

\bibitem[{{Wang} {et~al.}(2007){Wang}, {Mo}, \& {Jing}}]{Wang_et_al_2007}
{Wang}, H.~Y., {Mo}, H.~J., \& {Jing}, Y.~P. 2007, \mnras, 375, 633

\bibitem[{{Wechsler} {et~al.}(2002){Wechsler}, {Bullock}, {Primack},
  {Kravtsov}, \& {Dekel}}]{Wechsler_et_al_2002}
{Wechsler}, R.~H., {Bullock}, J.~S., {Primack}, J.~R., {Kravtsov}, A.~V., \&
  {Dekel}, A. 2002, \apj, 568, 52

\bibitem[{{Wechsler} \& {Tinker}(2018)}]{Wechsler_Tinker_2018}
{Wechsler}, R.~H., \& {Tinker}, J.~L. 2018, \araa, 56, 435

\bibitem[{{Wechsler} {et~al.}(2006){Wechsler}, {Zentner}, {Bullock},
  {Kravtsov}, \& {Allgood}}]{Wechsler_et_al_2006}
{Wechsler}, R.~H., {Zentner}, A.~R., {Bullock}, J.~S., {Kravtsov}, A.~V., \&
  {Allgood}, B. 2006, \apj, 652, 71

\bibitem[{{Wetzel} {et~al.}(2007){Wetzel}, {Cohn}, {White}, {Holz}, \&
  {Warren}}]{Wetzel_et_al_2007}
{Wetzel}, A.~R., {Cohn}, J.~D., {White}, M., {Holz}, D.~E., \& {Warren}, M.~S.
  2007, \apj, 656, 139

\bibitem[{{Wetzel} {et~al.}(2014){Wetzel}, {Tinker}, {Conroy}, \& {van den
  Bosch}}]{Wetzel_et_al_2014}
{Wetzel}, A.~R., {Tinker}, J.~L., {Conroy}, C., \& {van den Bosch}, F.~C. 2014,
  \mnras, 439, 2687

\bibitem[{{White} \& {Rees}(1978)}]{White_Rees_1978}
{White}, S.~D.~M., \& {Rees}, M.~J. 1978, \mnras, 183, 341

\bibitem[{{Zentner}(2007)}]{Zentner_2007}
{Zentner}, A.~R. 2007, International Journal of Modern Physics D, 16, 763

\bibitem[{{Zentner} {et~al.}(2014){Zentner}, {Hearin}, \& {van den
  Bosch}}]{Zentner_et_al_2014}
{Zentner}, A.~R., {Hearin}, A.~P., \& {van den Bosch}, F.~C. 2014, \mnras, 443,
  3044

\bibitem[{{Zhao} {et~al.}(2009){Zhao}, {Jing}, {Mo}, \&
  {B{\"o}rner}}]{Zhao_et_al_2009}
{Zhao}, D.~H., {Jing}, Y.~P., {Mo}, H.~J., \& {B{\"o}rner}, G. 2009, \apj, 707,
  354

\bibitem[{{Zhao} {et~al.}(2003){Zhao}, {Mo}, {Jing}, \&
  {B{\"o}rner}}]{Zhao_et_al_2003}
{Zhao}, D.~H., {Mo}, H.~J., {Jing}, Y.~P., \& {B{\"o}rner}, G. 2003, \mnras,
  339, 12

\end{thebibliography}

\appendix

\section{Effects of halo definition on concentration in the Rockstar halo finder}
\label{app:halo_definition_rockstar}

The \textsc{Rockstar} halo finder works by dividing the simulation into 3D friends-of-friends (FOF) groups, adaptively creating smaller 6D FOF groups in phase space, placing halo centres at the most refined 6D FOF groups, and finally calculating halo properties relative to those centres \citep{Behroozi_et_al_2013_1}.  The size of the initial 3D FOF groups is set by the input linking length in units of the mean interparticle separation, $b$.  The accuracy of the halo properties computed by \textsc{Rockstar}  depends on the original 3D FOF groups percolating out to the baseline overdensity radius of the corresponding haloes \citep{More_et_al_2011,Behroozi_et_al_2013_1}. This is particularly important when fitting halo density profiles: if halo boundaries extend into the unpercolated regions of the FOF group, the density in the outermost radial bins will be systematically underestimated, shifting the location of profile features. In this Appendix, we examine the effect of using different halo definitions on the measured concentrations.

\citet{Behroozi_et_al_2013_1} perform convergence tests that show that using linking length of $b=0.28$ leads to full percolation within $R_{\rm vir}$. They note that when one defines halo boundary larger than $R_{\rm vir}$, a larger linking length should be used and additional tests should be performed to ensure full percolation within such boundary. 
We test the effect of the halo boundary choice by running \textsc{Rockstar} repeatedly on the CBol\_L0125 simulation from \citet{Diemer_Kravtsov_2015} for a variety of overdensity radii, $R_\Delta$, with overdensity ranging from \,$\Delta = 20 \rho_{\rm m}$ to $\Delta=1600\,\rho_{\rm m}$ for $b=0.28,0.5$ and for a reference catalogue with $\Delta=\Delta_{\rm vir}$ and $b=0.28$. We then matched haloes across the catalogues to our reference $R_{\rm vir}$ catalogue. Our tests indicate that results are not sensitive to the way this matching is done, so we use a simple procedure where a halo is considered a ``match'' if its centre lies within 0.25 kpc of the centre of a counterpart in the $R_{\rm vir}$ catalogue. This criterion is sufficient to unambiguously match most haloes, but in the event that multiple haloes meet it, we match to the halo in that group with the closest $M_{\rm vir}$ to the reference halo. Subhalo and distinct halo status are not factored in to this matching. We restrict our sample to haloes classified as hosts by the  $\Delta_{\rm vir}$ catalogue with $10^{11.5}\,h^{-1}\,M_\odot< M_{\rm vir} < 10^{12}\,h^{-1}\,M_\odot,$ as measured by the same catalogue. The choice of mass range has only a slight effect on results. 

\begin{figure}
   \centering
   \includegraphics[width=\columnwidth]{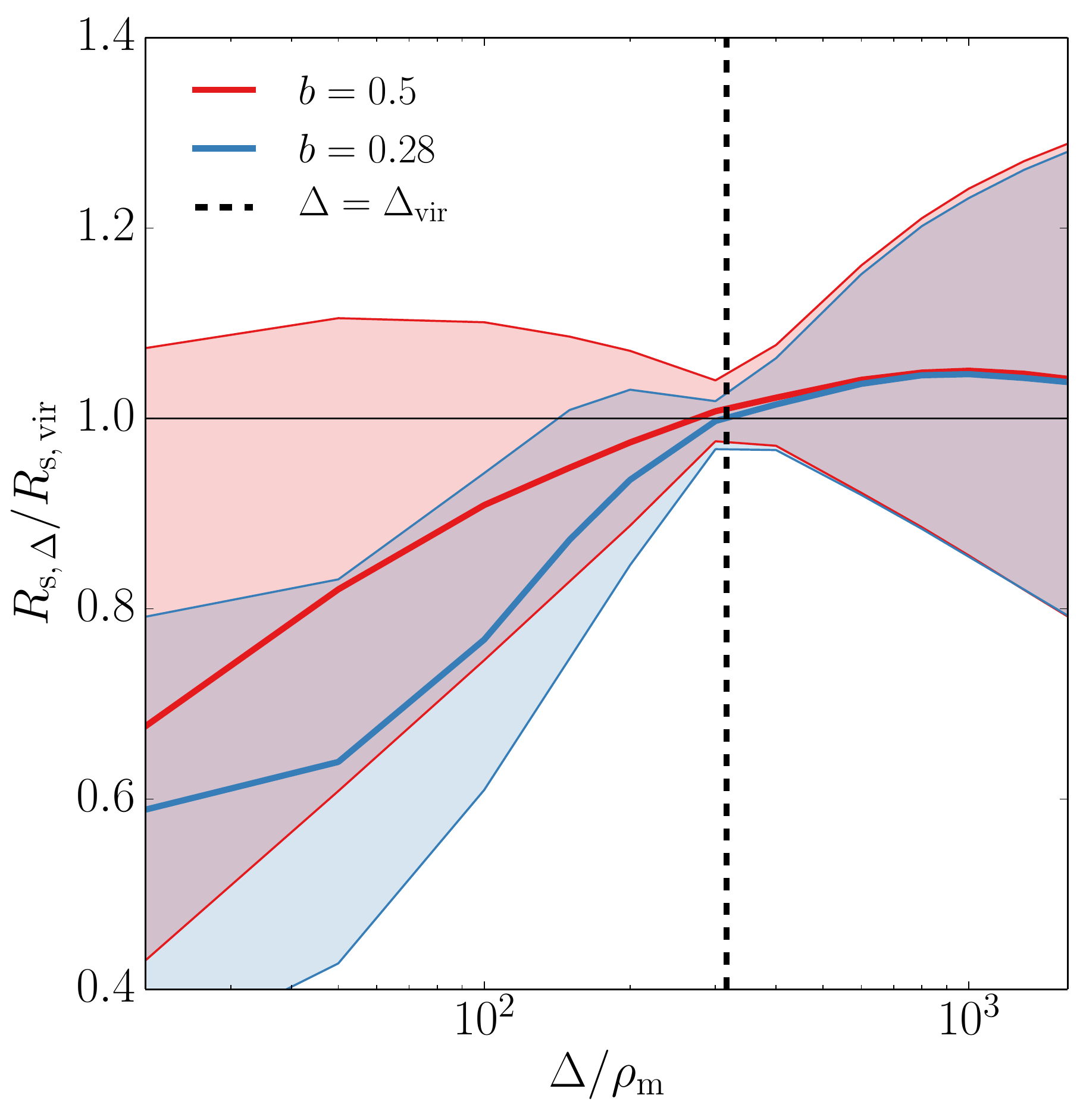}
   \caption{The value of $R_{\rm s}$ measured by \textsc{Rockstar} using different overdensities, $\Delta$, to define halo radius, $R_\Delta$. The scale radii are normalized by the value of $R_{\rm s}$ measured for the same haloes in a catalogue with a primary definition of $\Delta_{\rm vir}$ for two different values of the \textsc{Rockstar}'s 3D FOF linking length, $b$. The median values of this ratio are shown as solid lines and the contours enclosing 68\% of ratios are shown as shaded regions. Note that $R_{\rm s}$ measurements for $\Delta\lesssim 200$ are biased low relative to the values found for $\Delta = \Delta_{\rm vir}$ and there is a significant scatter between the two. The bias is larger for smaller $b$.}
  \label{fig:ratio_compare}
\end{figure}

Fig.~\ref{fig:ratio_compare} shows the ratio  of $R_{\rm s}$ measured in catalogues constructed for  different values of $\Delta$, denoted $R_{\rm s,\,\Delta},$ to $R_{\rm s}$ measured in the reference catalogue, denoted $R_{\rm s,\,vir}$. We show this ratio as a function of $\Delta$ for two values of $b$. We find that creating catalogues with larger linking lengths takes an inordinate amount of time, presumably because a large fraction of the simulation is placed into the same FOF group.  
The figure shows that .$R_{\rm s}$ measurements for $\Delta\lesssim 200$ are biased low relative to the values found for $\Delta = \Delta_{\rm vir}$ and there is a significant scatter between the two. The bias is about twice larger for $b=0.28$ compared to $b=0.5$.  

The primary implication of this result is that concentrations measured for haloes identified and analyzed by \textsc{Rockstar} using low $\Delta$, such as $\Delta=20\,\rho_{\rm m},$ should not be trusted due to large systematic bias and scatter. This is due to lack of FOF percolation in the outskirts of haloes, which biases densities in the outskirts low and this, in turn, biases the best-fit $R_{\rm s}$ values low. This is also true for other ways of estimating concentration, such as deriving it from $V_{\rm max}/V_{\rm vir}.$  The context of this fact in relation to our work on assembly bias is discussed \ref{ssec:comparison_with_previous_analysis}.

Behaviour at commonly used choices, such as $\Delta=200\,\rho_{\rm m},\,200\,\rho_{\rm c},\,500\,\rho_{\rm c}$ is also noteworthy. For $b=0.28,$ the systematic biases on $R_{\rm s}$ for these three definitions relative to our reference $\Delta=\Delta_{\rm vir}$ catalogue  are +6\%, -5\%, and -5\%, respectively. While the difference for $\Delta=200\rho_{\rm m}$ has contributions from lack of percolation, the difference between $\Delta_{\rm vir}$ and higher density definitions must be due to a different effect, such as deviations of halo profiles from the fitted NFW form. Any attempt to compare, for example, mass-concentration relations to the $\approx 5\%$ level measured with different primary definitions should account for this effect.

Lastly, as discussed in section \ref{ssec:subhaloes}, overdensity radii are fundamentally unphysical choices for halo boundaries, and $\Delta_{\rm vir}$ cannot be thought of as a more ``correct'' choice than other nearby overdensities. Consequently, Fig.~\ref{fig:ratio_compare} should not be interpreted as showing deviations from the true value of $R_{\rm s}$, but merely deviations from a particular reference value where the FOF groups are known to be percolated.

\section{Fast Halo Containment Checks}
\label{app:fast_halo_containment_checks}

Numerous components of the analysis presented in this paper rely on containment checks, particularly when computing subhalo status, computing $R_{\rm tidal}$, or computing $M_{\rm tidal}$. Out sample contains $\approx 300,000$ haloes and the Bolshoi simulations contain $2048^3$ particles each, so a naive $N^2$ check of every pair of objects would be prohibitively expensive. This is particularly true when identifying splashback subhaloes through the surfaces found by \textsc{Shellfish} because \textsc{Shellfish} represents splashback shells using third-order Penna-Dines surfaces, which take roughly fifty math library function calls to evaluate. In this Appendix, we describe our approach for computing containment checks.

First, suppose we are given a set of points $P$, which we must check for containment within a set of haloes, $H.$ First, we construct a uniform 3D grid spanning the simulation volume and place all elements of $P$ within lists associated with each cell in the grid. Then, for each halo in $H$ we construct a bounding box fully enclosing its boundary and compute containment checks for only the particles which reside in grid cells that intersect with it.
Because the lists associated with each grid cell are created once and potentially iterated over many times, we represent lists as dynamically allocated arrays instead of as linked lists to increase cache locality.  We find that for a grid with $250^3$ cells containment checks are no longer a significant component of the runtime cost of any analysis in this paper.

In the case where a halo boundary is determined by an expensive function $f(\phi, \theta)$, such as the Penna-Dines functions used  by the \textsc{Shellfish} code to approximate splashback shells, we use the following procedure to accelerate containment checks. First, for every halo in $H,$ we compute the minimum and maximum values of $f(\phi,\theta)$, $f_{\rm max}$ and $f_{\rm min}$. Since a point at a distance $r$ is automatically contained if $r<f_{\rm min}$ and automatically not contained if $r > f_{\rm max}$, we only evaluate $f(\phi,\theta)$ if the points is at a distance, $r$, that satisfies $f_{\rm min}<r<f_{\rm max}$.

\section{Tidal Force Errors}
\label{app:tidal_forces}

\begin{figure*}
    \centering
        \includegraphics[width=\columnwidth]{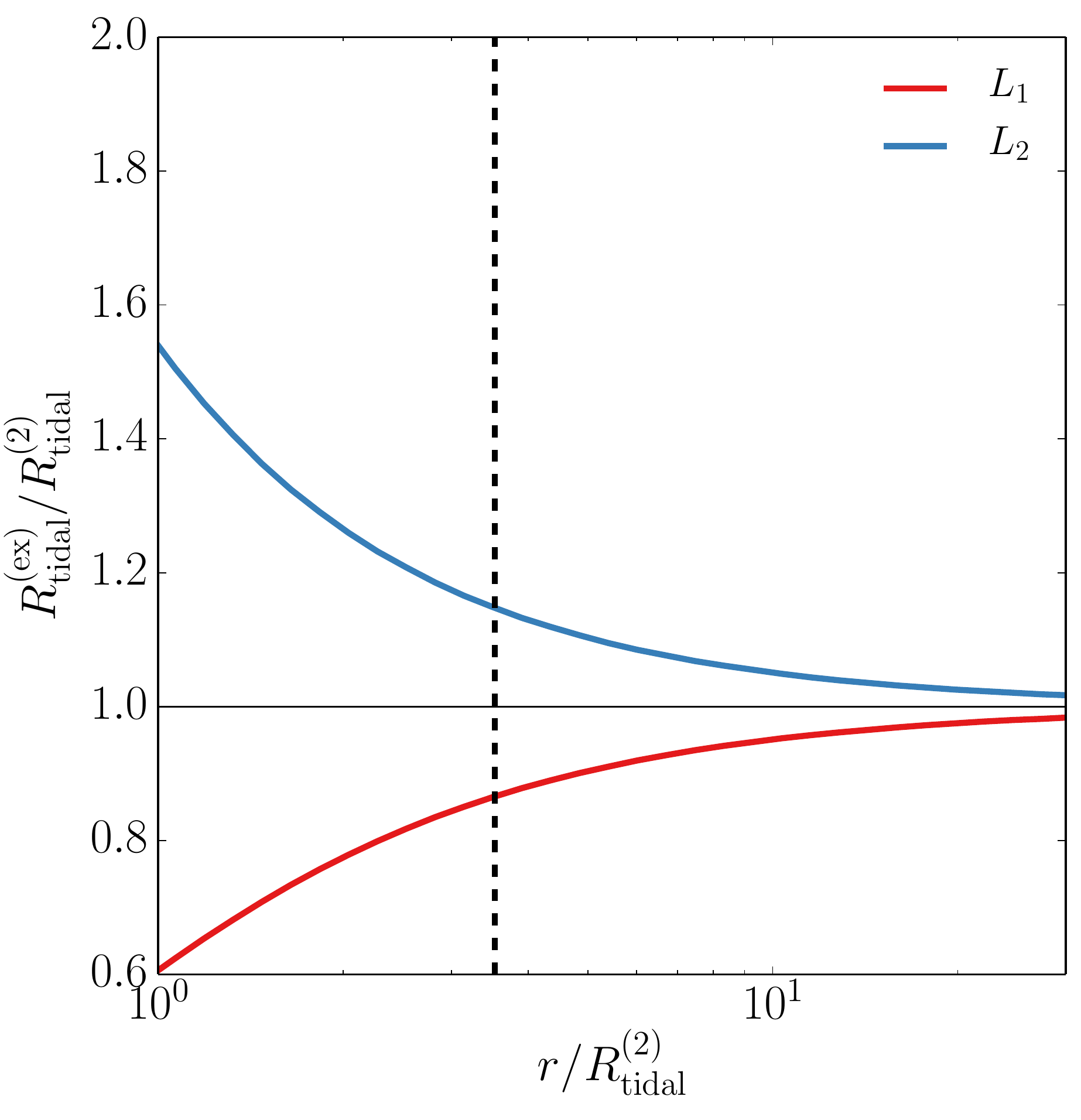}
        \includegraphics[width=\columnwidth]{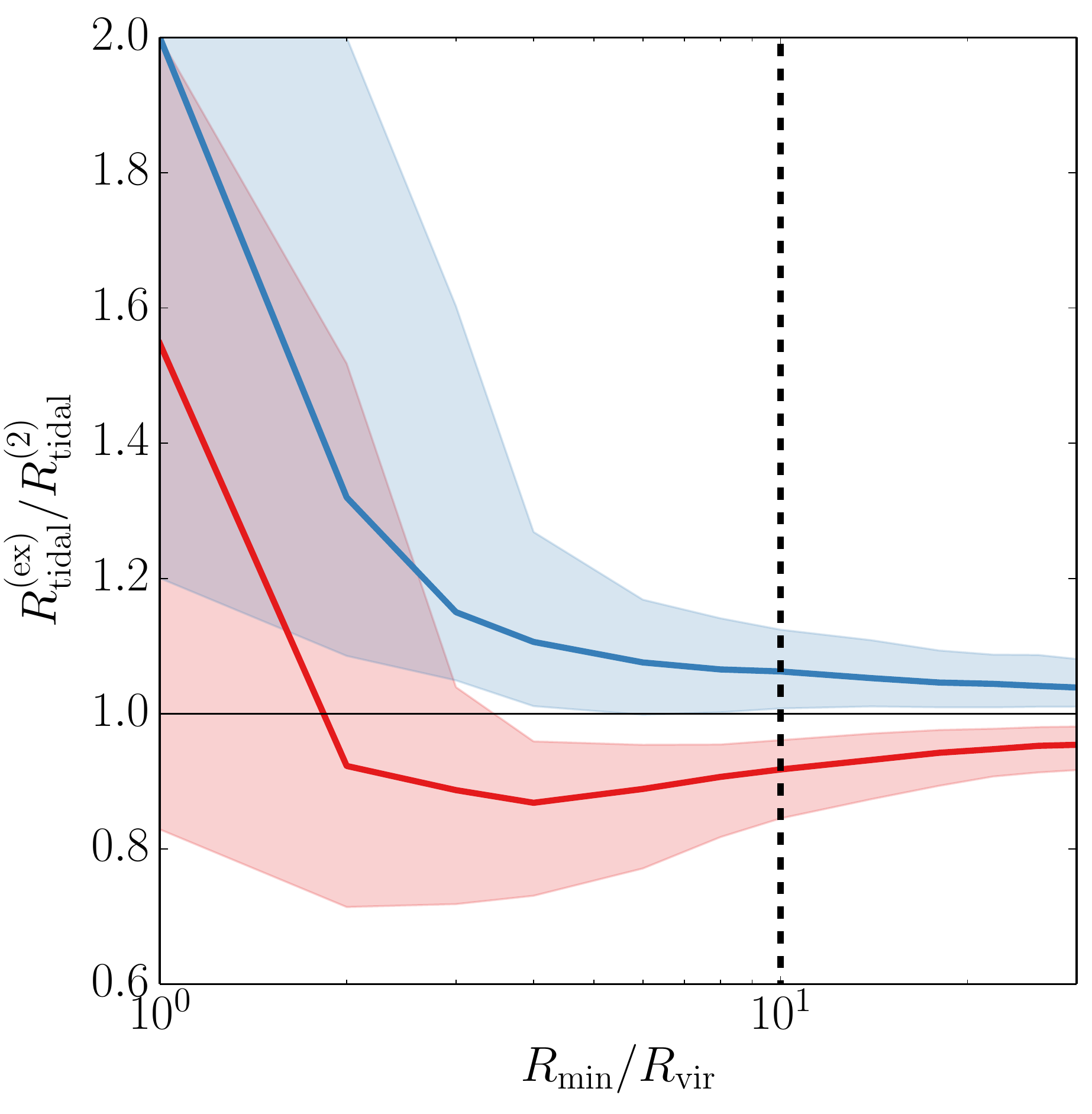}
    \caption
    {
        Two methods for estimating the error associated with different choices $R_{\rm min}.$
        Left: Analytic calculation of the error on $R_{\rm tidal}^{{2}}$ for a model system where the entire external potential is generated by a point source separated from the halo by a distance $r.$ The dashed black line shows the median value of $R_{\rm min}/R_{\rm tidal}^{(2)}$ for our sample. Right: The approximate error on $R_{\rm tidal}^{(2)}$ as a function of $R_{\rm min}/R_{\rm vir}$ for the haloes in our sample. Solid lines show the median values of $R_{\rm tidal}^{\rm (ex)}/R_{\rm tidal}^{(2)}$ and the shaded bands give the 68\% contours. $R_{\rm tidal}^{\rm (ex)}$ was estimated by evaluating $F_{\rm tidal}^{\rm (ex)}$ at $L_1^{(2)}$ and $L_2^{(2)}$ and applying the methods described in Appendix \ref{app:tidal_forces}. The black dashed line shows the value of $R_{\rm min}$ used in our analysis. Note that the x-axes of these two plots are scaled by different characteristic radii.
    }
    \label{fig:tidal_error}
\end{figure*}

In this Appendix, we investigate some of the error properties of the tidal radius and discuss an important approximation made in our calculation of $R_{\rm tidal}$, the inclusion of a minimum cutoff radius when adding contributions to a halo's local tidal tensor. For the purpose of clarity, we will refer to the tidal radius calculated after a second-order approximation of the external gravitational potential has been made as $R_{\rm tidal}^{(2)}$ and the tidal radius when the exact tidal field is used as $R_{\rm tidal}^{(\rm ex)}$. Other quantities will use an analogous referencing scheme. Elsewhere in this paper, $R_{\rm tidal}^{(2)}$ is referred to as  $R_{\rm tidal}$. We also take the convention that the Lagrangian point between a halo and an external source is $L_1$ and that the point on the opposite side of the halo is $L_2.$ In cases where analysis is performed on haloes without a single external source, $L_1$ is the Lagrangian point with the lowest external potential.

Like the classical $R_{\rm Hill}$ derivation, our calculation of $R_{\rm tidal}^{(2)}$ (see Equation \ref{eq:r_tidal}) assumes that the external tidal field felt by the halo is well-approximated by a second-order hyper-paraboloid. This is necessary because the tidal tensor which is used to determine the principle components of the tidal field only contains second derivatives of the gravitational potential. Note that in the special case where there is only a single external point source,
\begin{equation}
	\label{eq:tidal_to_hill}
	R_{\rm tidal}^{(2)} = \left(\frac{3}{2}\right)^{1/3} R_{\rm Hill}^{(2)},
\end{equation}
so the discussion below can be extended to error analysis on the classical Hill radius. Note that the factor of $(3/2)^{1/3}$ is because the derivation of $R^{(2)}_{\rm Hill}$ assumes that the halo is on a circular orbit around the external point source and thus experiences a centrifugal force in addition to a tidal force, while the derivation of $R_{\rm tidal}^{(2)}$ assumes that all non-tidal pseudo-forces are zero. Given the scale of errors discussed below, and the fact that this factor decreases as haloes deviate from circular orbits -- a configuration which is very rare for distinct haloes -- we do not consider this difference to be significant.

We perform two complementary tests on the accuracy of $R_{\rm tidal}^{(2)}.$ First, we analytically compute $R_{\rm tidal}^{(2)}/R_{\rm tidal}^{(\rm ex)}$ for a single source at a distance $r$ from a halo, and second, we measure exact the error on $F_{\rm tidal}^{(2)}$ for our halo sample and combine this with reasonable assumptions about the shape of the tidal field to estimate upper limits on $R_{\rm tidal}^{(2)}/R_{\rm tidal}^{(\rm ex)}.$

We show the results of this first calculation in the left panel of Fig.~\ref{fig:tidal_error}. Although there is no closed-form expression for $R_{\rm tidal}^{(\rm ex)},$ it can be found numerically by maximizing the effective potential. We parameterize the error as a function of $r/R_{\rm tidal}^{(2)}$ which also absorbs the dependence on the mass ratio. We recover the well-known fact that as the mass ratio between the halo and the external source decreases and the tidal radius increases, the two Lagrangian points become asymmetric and that errors become increasingly significant. This can also be interpreted as an estimate of the error associated with a particular value inner cutoff radius for $r=R_{\rm min}$ when following the procedure described in section \ref{ssec:r_tidal}. This can be considered a worst-case estimate of the error at a given $R_{\rm tidal}^{(2)}$ because the true matter distribution will generally contain many points at distances larger than $R_{\rm min}$ which contribute significantly to the tidal field.

We perform our second test by first computing $L_1^{(2)},$ $L_2^{(2)},$ $R_{\rm tidal}^{(2)},$ and $F_{\rm tidal}^{(2)}$ for every halo in our sample for a particular choice of $R_{\rm min}.$ Then, we use the raw particle data to compute the radial and tangential components of $F_{\rm tidal}^{(\rm ex)}(L_1^{(2)})$ and $F_{\rm tidal}^{(\rm ex)}(L_2^{(2)})$ for these haloes. Particles within $R_{\rm min}$ are not included in this calculation. To obtain an estimate of $R_{\rm tidal}^{(\rm ex)}$ from this, we make two simplifying assumptions about the the shape of the tidal field. First, we assume that the exact Lagrangian points lie along the same axis as a halo's second-order Lagrangian points. We find that the tangential components of $F_{\rm tidal}^{(\rm ex)}(L_1^{(2)})$ and $F_{\rm tidal}^{(\rm ex)}(L_2^{(2)})$ are small compared to the radial components, implying that this is a reasonable assumption. Second, we assume that along the lines connecting $L_1^{(2)}$ to $L_1^{(\rm ex)}$ and $L_2^{(2)}$ to $L_2^{(\rm ex)},$ the tidal force varies slowly enough that it can be well approximated by
\begin{equation}
    F_{\rm tidal}^{\rm (ex)}(r) \propto \left(\frac{r}{R_{\rm tidal}^{(2)}}\right)^{1+\alpha},
\end{equation}
Here, $\alpha$ is an arbitrary constant which varies from halo to halo and may be different for different Lagrangian points within the same halo. It represents the deviation from the scaling seen when the tidal potential is approximated to second order. In this case, 
\begin{equation}
    \frac{R_{{\rm tidal}}^{(\rm ex)}}{R_{\rm tidal}^{(2)}} =
    \left(\frac{F_{\rm tidal}^{(\rm ex)}(L_i^{(2)})}{F_{\rm tidal}^{(2)}(L_i^{(2)})}\right)^{-\frac{1}{3+\alpha}},
\end{equation}
where $i$ indexes over Lagrangian points. This assumption is informed by tests on single-source effective potentials, which find that for all but the smallest external point sources, $-1<\alpha<0$. For haloes where $R_{\rm min} > R_{\rm tidal}^{(2)},$ we would expect that at a constant $R_{\rm tidal}^{(2)}$ the tidal field would be varying more quickly at $L_i^{(2)}$ when the field is generated by a single point source than when it is generated by a more diffuse matter distribution, so it's likely that this range of $\alpha$ values holds for our simulated haloes as well. For this reason we can place the following upper bound on the error in tidal radius:
\begin{equation}
    |R_{\rm tidal}^{(\rm ex)}/R_{\rm tidal}^{(2)} - 1| < |F_{\rm tidal}^{(\rm ex)}/F_{\rm tidal}^{(\rm 2)} - 1|^{-1/2}.
\end{equation}

We show the fractional error in $R_{\rm tidal}$ using this limit in the right panel of Fig.~\ref{fig:tidal_error} as a function of the adopted $R_{\rm min}.$ Errors balloon uncontrollably for $R_{\rm min} \lesssim 4\,R_{\rm vir}$ but are more well-behaved at larger radii, with errors dropping to the $\approx 10\%$ level at $\approx 10\,R_{\rm vir}.$

One interesting feature of this Figure is that for $R_{\rm min} \lesssim 2\,R_{\rm vir},$ the error on the location $L_1$ becomes positive. This is likely because this is the characteristic size of the splashback radius, meaning that the halo's own particles will be incorporated into the calculation of $F_{\rm tidal}^{(\rm ex)}.$ Since the tidal force is repulsive, this inclusion of halo particles will reduce the apparent strength of the field and increase $R_{\rm tidal}.$

These tests indicate that for $R_{\rm min} = 10\,R_{\rm vir},$ the errors in $R_{\rm tidal}^{(2)}$ which are specifically due to the second order approximation of the tidal field are small. However, this analysis is performed at a constant $R_{\rm min},$ so it doesn't account for errors due to the removal of significant sources close to the halo. This is not an issue for our analysis because $R_{\rm tidal}$ is explicitly a proxy for the \emph{large-scale} tidal field, and our proxy $D_{\rm vir}$ is better suited for close sources. This would, however, become a significant issue for studies which need $R_{\rm tidal}$ for purposes other than rank-ordering haloes. Further discussion on the impact that improvements in the accuracy of $R_{\rm tidal}$ would have on our results can be found in section \ref{ssec:proxy_def}.

More generally, while the issue of measuring tidal radii around haloes with only a single significant source is well-explored \citep[see \S 2 in][for a review]{van_den_Bosch_et_al_2018}, and the tidal radius due to the large scale field can be measured effectively with the tidal tensor, there currently does not exist an effective method for combining these two regimes. We outline a number of potential approaches which could be used to address this issue in \ref{ssec:directions_for_future_work}, but consider the testing and calibration of such methods to be beyond the scope of this paper.

\section{Identifying Bound Particles in Halo Outskirts}
\label{app:boundedness}

While the concept of gravitational binding is straightforward to define for particles near the center of a non-accelerating halo, the same is not true for particles in the outskirts of haloes, especially those experiencing a strong tidal force. These difficulties arise from two key areas: first, it is difficult to disentangle the potential caused by a halo from the potential of its surroundings. Although haloes have a non-trivial amount of mass stored outside $R_{\rm vir},$ the so-called ``two-halo'' term starts to dominate the density distribution at $r\gtrsim 1 - 2 R_{\rm vir}$ \citep[e.g.][]{Diemer_Kravtsov_2014}, meaning that any calculation of the potential which is done directly from the density profile or from the particle distribution must be done with care. Second, for particles near the tidal radius, the effective potential due to the external tidal field becomes significant. While this issue could in principle be solved by defining escape velocities relative to the minimum potential at either Lagrangian point, it also means that boundedness calculations will suffer from the same accuracy issues as the tidal radius calculations (see Appendix \ref{app:tidal_forces}).

The effect of tidal forces on particle escape velocities presents another issue for the analysis in this paper, specifically. If this effect is taken into account, it means that \emph{gravitational heating and tidal forces can no longer be disentangled}. Even a ``control'' variable like $M_{\rm \beta,b}$ would depend on the tidal field, and could potentially make gravitational heating appear to be more a more significant contributor to assembly bias than it actually is. Primarily because of this reason, and to a lesser extent because of the issues described in the previous paragraph, we take on the simple and standard boundedness condition given in Eq.~\ref{eq:v_esc}, but note that the $f_{\rm removed}$ value for $M_{\rm tidal,b}$ could become even lower if more sophisticated approaches were used.

\label{lastpage}

\end{document}